# Political Hegemony, Imitation Isomorphism, and Project Familiarity: Instrumental Variables to Understand Funding Impact on Scholar Performance

**Yang Ding (0000-0002-0683-2655)** [a,b,c] **Yi Bu (0000-0003-2549-4580)** [d,e,f,*]

[a] Business School, University of Edinburgh, Edinburgh EH8 9JS, United Kingdom

[b] Lab for Interdisciplinary Spatial Analysis, University of Cambridge, Cambridge CB3 9EP, United Kingdom

[c] Department of Land Economy, University of Cambridge, Cambridge CB2 1RX, United Kingdom

[d] Department of Information Management, Peking University, Beijing 100871, China

[e] Center for Informationalization and Information Management Research, Peking University, Beijing 100871, China

[f] Center for Digital Intelligence Science and Education Research, Peking University Chongqing Research Institute of Big Data, Chongqing 401332, China

* Corresponding Author: Yi Bu (buyi@pku.edu.cn).

**Abstract**

This paper contributes a new idea for exploring research funding effects on scholar performance. By collecting details of 9,501 research grants received by principal investigators from universities in the U.S. social sciences from 2000 to 2019 and data on their publications and citations in the Microsoft Academic Graph and Web of Science bibliographic collections, we build a novel dataset of grants and article counts, citations, and journal CiteScore. Based on this dataset, we first introduce three instrumental variables (IVs) suitable for isolating endogeneity issues in the study of competing grant effects, namely scholars' political hegemony in academia, imitation isomorphic behavior among scholars, and project familiarity. Then, this study explains the research funding effects by combining the three IVs with a two-stage least square (2SLS) model. Also, we provide validity and robustness tests of these three IVs and research funding effects. We find that our IVs serve the function of exogenizing and isolating endogeneity in capturing the research funding effect. Empirical findings show that receiving research funding increases a scholar's research output and impact. While research funding doesn't significantly increase high CiteScore publications, it reduces submissions to low-prestige journals, reshaping journal selection strategies and raising the "floor" of academic performance.

## 1. Introduction

In the winter of 1978, the outbreak of the Islamic Revolution in Iran resulted in an oil crisis that spread from Tehran to the rest of the world, causing inflation and economic depression in a wide range of countries including the United States. A few months later, Franco Modigliani from MIT became one of the first social and economic scholars to receive funding from the National Science Foundation, a $150,000 grant to assist governments and the scientific community in understanding the monetary mechanism and stabilization policy during the crisis. In the subsequent decades, nations have kept investing in social sciences with the expectation that their research will contribute to economic growth and societal advancement. Nonetheless, the public endlessly debates whether investment in academia will eventually contribute to the discovery and dissemination of knowledge. Early in the 20th century, the development of electronic and open access movements (Evans & Reimer, 2009) in the literature increased the possibility of quantifying and assessing the productivity of science, necessitating the theorization and investigation of the beyond correlation relationship between research funding and scientific productivity.

In practice, the role of research funding in the day-to-day research of academics is being debated. Generally, most scholars claim that investment in science affects scientific labor (Crespi & Geuna, 2008). In other words, research funding is expected to free scientists from financial issues, equipment shortages, and academic communication challenges (Zhang et al., 2024), allowing them to focus more on their intellectual work and affecting their knowledge discovery and innovation progress. Nevertheless, as research funding has become increasingly competitive, it may be having a negative effect on academics (Alkhawtani et al., 2020; Carvan, 2022). For example, the considerable time and effort required by academics to prepare proposals and applications (Bollen et al., 2014), with a high risk of not receiving funding, may be dragging down their output. This endless competition and the negative consequences make some researchers resistant to applying for grants. Further, even if they are funded, academics may be trapped with a large number of teaching responsibilities or household duties (Malisch et al., 2020) while researching and reporting regularly on their progress, exacerbating the stress of their work. Thus, it remains to be determined whether the role of research funding is what governments expect it to be.

Yet, extant empirical results provide controversial conclusions about the relationship between research funding and scholars' performance. Detailed evidence of the relationship between funding and academic performance, and the controversies between such evidence, is provided in **" A. MOTIVATION" of the Supplementary material**. This controversial conclusion is reflected in whether research funding is effective in terms of scientific output and, if so, the directionality of this impact, i.e. positive or negative (Alkhawtani et al., 2020; Beaudry & Larivière, 2016; Benavente et al., 2012; Boyack & Jordan, 2011; Ebadi & Schiffauerova, 2016; Hottenrott & Thorwarth, 2011; Jacob & Lefgren, 2011; Leydesdorff et al., 2019; Pagel & Hudetz, 2015; Sandstrm & van den Besselaar, 2018; Wagner et al., 2018; Wang et al., 2019). The reasons for these controversies could be considerably complicated, partly because it is not yet clear which is the cause and which is the effect of research funding and scholar performance. Research funding decisions are routinely driven by the applicant's past productivity or allocated to scholars/and projects with greater potential (Van Arensbergen & Van Den Besselaar, 2012), leading to a reverse causal relationship between research funding and scholar performance. Then, research funding may further proliferate the scientific publications they are funded for, further leading to selection bias (Ebadi & Schiffauerova, 2016; Kaiser, 2019; D. Li et al., 2017). Thus, the selection bias and potential reverse causality resulting from such biased and selective allocation rules for research funding make it possible to confound research funding effects when comparing funded and non-funded scientists. For example, some studies confirm that these endogeneity issues such as selection bias can be expressed as potential, inherent productivity gaps between individual scholars (W. Li et al., 2022), or publishing discrepancies due to gaps in the academics' research areas and affiliations (Shin et al., 2022).

Recent evidence suggests that scholars expect to use beyond correlation strategies to capture research funding's impact on scholar performance. Instrumental variables (IVs) are attracting much attention because of their strength in isolating endogeneity issues caused by research funding. For example, Yu, Dong, and de Jong (2022) use the urban economic dynamism of universities as an IV to isolate the potential endogeneity of research investment in China when assessing the impact of public research funding on the research output and quality of 622 universities in China from 2010-2017. Yet, it is invariably challenging to find exogenous factors that do not directly affect scholar performance but rather indirectly affect scholar performance by shaping the probability of scholars receiving funding, resulting in a rather limited selection of currently available IVs (Andrews et al., 2019). Moreover, most IVs measuring funding effects are concentrated in non-English speaking countries (Lawson et al., 2021). Therefore, it is imperative to find or design IVs for research grants in English-speaking countries wherever possible to isolate selection bias, omitted variables, etc. and to ensure cleaner grant validity outcomes.

This paper examines the impact of receiving research funding on scholars' performance, focusing on competitive funding in English-speaking nations. Using the Social, Behavioral and Economic Sciences (SBE), one of the National Science Foundation's (NSF's) directorate, grants from 2000 to 2019 as examples, we adopt IV estimations and two-stage least square (2SLS) regressions to isolate and understand the aforementioned endogeneity of research funding that confounds the assessment of effects. These enable us to quantify the effect of research funding on the number of publications and the impact of studies. Specifically, the main contribution of this paper is that we present and introduce three IVs in research funding, namely scholars' political resources, isomorphic behavior, and scholars' familiarity with the SBE program application. We explain how we construct these three IVs and provide tests of their validity and robustness, expecting other scholars to consider them when estimating treatment effects related to academic performance. Lawson et al. (2021) examined the utility of the first IV in funding competitions in non-English-speaking nations. We introduce it to English-speaking countries and combine it with the second and third IVs to demonstrate their efficacy in isolating the endogeneity of research funding. As far as we know, two IVs, isomorphic behavior and scholars' familiarity with program application, are proposed for the first time. Simultaneously, unlike most studies that use complex and varying competing grants as a treatment, this study relies on the most common standard and continuing grants in the US social sciences to further control for possible confounding of grant effectiveness assessment by multi-purpose types of grants. We take into account the differences of initial productivity, gender, institutions, and research fields between scholars, which further reduces the differences between samples caused by reverse causality and missing variables. Empirical evidence suggests that SBE research funding improves the quantity and quality of academics' research significantly, but hardly causes their work to be published in more prestigious journals.

## 2. Theoretical backgrounds

IVs are considered by many scholars as one of the most effective strategies to address endogeneity issues. Nevertheless, the exogenous requirements of IVs means that their options for exploring the relationship between research funding and scholar performance are fairly limited. By summarizing some of the existing research, we find that the available IVs broadly fall into two categories: lagged variables and exogenous variables designed on the basis of 'local conditions' (A. N. Berger et al., 2005). The former is more common in empirical evidence in economics and business, and the core idea is to use lags of endogenous variables as IVs wherever possible. For instance, when measuring the impact of research funding on research output and innovation in 24 public universities in Ethiopia between 2016 and 2020, Abibo, Muchie, Sime, and Ezezew (2022) use all available lags of research funding as weak instruments. Similarly, Nugent, Chan, and Dulleck (2022) employ lagged values of scholars' patent activity and grants as IVs to explore the scholarly performance of 2,375 scholars at 36 Australian universities after receiving ARC funding. However, some scholars also express concerns about using such IVs. When the omitted variable that leads to endogeneity is autocorrelated, using the lagged endogenous variable as an IV does not result in consistent estimates of coefficients (Bellemare et al., 2017).

Further, some scholars use instruments based on local conditions. Kelchtermans, Neicu, and Veugelers (2022) design the award number of funded colleagues as an IV to estimate research funding effects on the academic performance of 734 scholars at KU Leuven, Belgium. They argue that the number of grants awarded to other academics around the target scholar in the year in which the target scholar received the grant implies the degree of desire or demand for research funding for the university, i.e., local resource scarcity. Political resources of scholars, funding availability of laboratories and economic indicators of university locations are also proposed to address the dilemma of having no available IVs other than lagged endogenous variables (Lawson et al., 2021; Yu et al., 2022).

## 2.1 Political hegemony

The scholars' political resources are frequently seen as an add-on to their ability to obtain more funding resources. When the effectiveness of the evaluation system depends on the decisions of individual reviewers, the authority to allocate resources may be transferred selectively and inexpressively in the form of grants to those specific individuals preferred by the evaluators (Braun, 1998; Groeneveld et al., 1975). It may be confusing to quantify the impact of such allocation mechanisms on fairness when exposed to human bias. Yet, there is an argument that internal endorsement by elite groups and cronyism have led, for example, to grants going to certain scholars with political resources, particularly in highly competitive environments for academic resources (Viner et al., 2004). Evidence suggests that academics employed as NSF rotators have access to more individuals outside of academia and receive twice as many research resources such as grants (approximately $200,000) compared to their colleagues in the same department over a period of five years (Hoenen & Kolympiris, 2020).

Despite the effect of political resources on scholars' resources, political resources may be controversial as an IV for capturing funding effects on scholarly performance. On the one hand, some studies claim that scholars' political resources can affect not only the probability of scholars receiving grants, but also their publication efficiency and research quality (Liu & Zhou, 2022), i.e., political resources do not fit the exogenous hypothesis of an IV. On the other hand. The political resources of scholars may be a rather vague definition. In general, scholars' political resources could be split by some studies into internal at the research institution and external tenure. For example, an internal administrative position within a university is considered by some scholars to be an academic's political resources. Political resources are claimed to be external to scholars when they hold senior positions in disciplinary associations off-campus or when they serve in leadership positions in government or other non-profit organizations. Political resources granted to scholars by different governments or funding agencies may manifest themselves differently for scholars (Viner et al., 2004). Therefore, when focusing on particular research issues, political resources of scholars should be further specified. Blind faith in or exclusion of political resources as an IV for scholars' academic resources or academic performance may not be based on solid evidence. For example, Cattaneo, Horta, and Meoli (2018) finds that academics receive dual appointments at universities and government or non-profit institutions and that such dual appointments do not have a significant effect on the academic performance of scholars. The availability of political resources as an IV is also shown to exist in specific countries by Lawson et al. (2021). The validity of political resources as an IV should therefore be further tested in practice based on the target.

## 2.2 Imitation isomorphism

Imitation isomorphism is defined as the tendency for an individual or organization to imitate another successful individual or organization when faced with uncertain goals/tasks because the former believes that some of the latter's behaviors are beneficial. Imitation isomorphism may be used as a potential exogenous variable based more on local conditions. individual scholars are frequently affiliated with a faculty or department in a university (P. Berger & Luckmann, 2016). This supports a cultural view that physical organizations often internally need to group individuals to meet the needs of individual roles and related tasks required for organizational development. This is particularly true of organizations and faculty settings within universities (Desimone, 2002). When similar goals such as promotion or resource needs exist between academics within a sub-group within a university, scholars may tend to imitate the successful scholars around them.

To clarify the direct relationship between imitation isomorphism and research funding, we can draw upon established theories of organizational behavior and social learning. Rooted in institutional theory, imitation isomorphism suggests that in resource-constrained environments, individuals are inclined to mimic behaviors they perceive as conferring a competitive advantage (DiMaggio & Powell, 2000). This process is not merely heuristic; it is a recognized pathway for navigating complex academic systems (DiMaggio & Powell, 2000; March et al., 1976). In the highly competitive landscape of academic funding, scholars often model their approaches on the successful strategies of colleagues within the same institution and research fields (Antonelli & Crespi, 2013). This aligns with organizational sociology's assertion that such imitation can lead to the emergence of similar research agendas and funding strategies within academic departments. Moreover, imitation isomorphism can directly impact scholars' research funding by facilitating the flow of information within academic communities (Lieberman & Asaba, 2006). As scholars observe their peers successfully obtaining funding, they gain insights into effective grant-writing techniques, funding opportunities, and strategic alignment with funding agencies' priorities (Velarde, 2018). This may enhance information exchange not only reduces the uncertainty associated with the funding process but also empowers scholars to make informed decisions that increase their likelihood of securing funding (Ordanini et al., 2008; Wood et al., 1992).

Furthermore, the theory of cumulative advantage, often referred to as the Matthew effect, underscores this relationship by highlighting how initial success in obtaining funding can create a self-reinforcing cycle. Scholars who secure funding early on are more likely to continue receiving it, fostering a pattern of success that others may seek to replicate (Bol et al., 2018; Merton, 1968). Consequently, scholars observing this pattern may be encouraged to pursue similar research areas or grant application strategies, thereby increasing their chances of securing funding through established channels. This dynamic is further supported by recent research demonstrating that scientific funding often flows to particular organizations and individuals, creating a snowball effect (Mitchell & McCambridge, 2022). When a university consistently receives funding for a specific research topic, it is more likely to maintain dominance in that area, which reflects a process of preferential attachment or the Matthew effect in research funding (Antonelli & Crespi, 2013).

Further, when imitation isomorphic trends occur, diffusion may lead to a cumulative effect (Seyfried et al., 2019). It has been shown that scientific funding is flowing to particular organizations and individuals and creating a snowball effect (Mitchell & McCambridge, 2022). When a university is regularly funded on a particular research topic, it has a greater tendency to establish dominance in that topic; it is then more likely to be funded on that topic later, in what is seen as a process of preferential attachment or the Matthew effect of research funding (Antonelli & Crespi, 2013). When isomorphisms among scholars develop, the grants of earlier winners do not seem to directly affect the scientific productivity of other scholars. Some scholars develop new IVs in an imitation isomorphic framework. For example, Kelchtermans et al. (2022) consider the number of research grants received by an academic's colleagues within the unit for the year as an IV, implying the demand for funding in the institution.

## 2.3 Project familiarity

Familiarity may affect the possibility of scholars receiving research grants, along two general pathways. The first path occurs at the wait-and-see stage of a scholar's grant application. How to choose the right foundation among a wide range of options and the setting of priorities is usually related to the scholar's cognitive needs (Ashrafian et al., 2010; Grass et al., 2017). Scholars' subjective cognitive needs may affect their pre-preparation and decision to apply for grants. Ashrafian et al. (2010) emphasize that scholars' cognition of grant conditions and normativity should be based on those grant opportunities with which they are more familiar, to increase the success probability.

The second path is reflected in the preparation of application materials, which can be understood as a game between the applicant and the funding decision maker. On the one hand, the accountability of funding agencies requires them to establish restrictive rules for research grants to initially guarantee the quality of the proposals they receive, allowing non-compliant proposals to slip out of the research grant application pipeline early. On the other hand, an applicant's familiarity with the annual grant topics and accuracy of the application process tends to affect the probability of their application being successful (van den Besselaar & Mom, 2022). In particular, in a more competitive funding environment where resources are scarce, applicants' proposals generally tend to be of high research merit or innovation. Therefore, in the evaluation of similar research programs, an applicant's understanding of the research projects and the application process and documentation is often one of the key reasons why that applicant stands out from the rest (Beleiu et al., 2015).

Despite the efforts of governments and funding agencies to communicate research objectives and guidelines through various channels, scholars remain largely unfamiliar with research funding processes. To address this gap, official NSF training sessions and workshops, which are highly visual and interactive, have become essential tools for enhancing scholars' understanding of funding opportunities [1]. These programs can significantly affect the likelihood of grant applications, whereas traditional announcements and documents often fall short. Moreover, there are distinct differences between the training related to writing research proposals and that for academic journal articles. Research proposals typically focus on crafting compelling narratives that align with funders' goals, while academic articles emphasize precise presentation of research findings and methodological rigor. According to a survey by Nature Masterclasses, a significant number of scholars encounter difficulties during the funding application process [2]. This underscores the importance of project familiarity as an exogenous factor influencing the link between research funding and academic performance.

3. **Data**

The scientific funding data comes from the SBE/NSF database. A detailed description of the SBE appears in **"B. SOCIAL, BEHAVIOURAL AND ECONOMIC RESEARCH FUNDING (SBE)" of the Supplementary material**. We keep information on scientific grants from standard and continuing programs and exclude grant types such as charity, student scholarships, and cooperative agreements. The data contains details of 9,501 U.S. university researchers who received SBE funding in the U.S. between 2000 and 2019. These researchers are supported as principal investigators (PIs) in the SBE's Social and Economic Sciences (SES), Behavioral and Cognitive Sciences (BCS), and SBE Office of Multidisciplinary Activities (SMA) sub-departments. We obtain the full names of these researchers, the funding amount, the start/end date of the grant, the type of grant, the grant number, the PIs' affiliation, the grant title, and the abstract of the grant. Moreover, we identify scholars' gender through Gender-API, and the identification method and results are shown in **"C. GENDER IDENTIFICATION BASED ON THE NAME OF THE PRINCIPAL INVESTIGATORS" of the Supplementary material**.

---

[1] https://beta.nsf.gov/events/join-us-nsf-day-university-wyoming-laramie
[2] https://masterclasses.nature.com/persuasive-grant-writing/20003200

The project names open for application by the SBE are diverse and constantly updated and added to over the years, making it difficult to observe the research topics of the funded projects to which academics belong in a straightforward way. Therefore, we allocate each academic to a research topic using the Latent Dirichlet Allocation (LDA) model (Blei et al., 2012) to extract topics from the titles and abstracts of all academics' programs. We obtained a total of 30 SBE-funded research topics, each of which included five keywords. The results of the LDA study topic sampling and the distribution of the project numbers are shown in **"D. TOPIC EXTRACTION OF FUNDED PROGRAMS THROUGH THE LDA MODEL" of the Supplementary material**. Descriptive metadata of 359,814 scientific publications by these academics in both the Web of Science (WoS) and Microsoft Academic Graph (MAG) databases is searched, covering the years 1995 to 2019.[3] Publications included in WoS provided more detailed grant information than other databases, allowing us to attribute grants to authors and publications more accurately. However, some studies confirm that the publications and journals included in WoS are incomplete compared to MAG and Scopus, resulting in a possible underestimation of author output and impact (Martín-Martín et al., 2021). Therefore, we conduct a joint search of WoS and MAG. We take the concatenation of the publications and highest value of citations for that recipient in the two databases. Our process for linking the SBE, WoS and MAG is shown in **"E. JOINT SEARCHING OF WEB OF SCIENCE (WOS) AND MICROSOFT ACADEMIC GRAPH (MAG) BIBLIOGRAPHIC DATABASES" of the Supplementary material**.

To measure the citation impact of a publication, we also obtain the CiteScore for the journal to which it belongs. CiteScore was designed and launched by Elsevier in 2016 to describe the performance of academic journals relative to other journals in the research area to which they belong.[4] While citation counts are frequently used as raw data to reflect the dissemination or popularity of a publication, CiteScore is seen as the citation impact of publications, and is often associated with academic prestige (DeJong & St. George, 2018). CiteScore is calculated by dividing the citation counts by the publication counts published in a given journal, over a four-year time window. It covers over 22,800 journals, more than twice the number of journals covered by another metric, the Journal Impact Factor (JIF). And, it solves the problem of JIF not being able to calculate citing records and cited lists of inaccessible articles, and is considered to be more transparent and accessible in its calculation (Fernandez-Llimos, 2018).

---

[3] This time window is selected because our grant data covers the years 2000 to 2019 and we need more than five years (1995-1999) to observe the historical performance of the recipients in the year 2000 in the years prior to receiving the grant.
[4] https://www.elsevier.com/solutions/scopus/how-scopus-works/metrics/citescore

As all these PIs come from U.S. universities, we add these universities to the U.S. News and World Report rankings. The U.S. News and World Report rankings focus more on teaching quality and faculty resources. Given the significant correlation between university reputation and the output of academics (Viner et al., 2004; Zhai et al., 2024), we also consider the QS university rankings which emphasize academic prestige and their university reputation scores. We use the ranking order of the universities in the 2022 rankings.[5]

Table 1 presents the statistical results of the variables included in the dataset, covering variables related to academics, research areas, affiliations, etc. At the individual scholar level, we can see that it takes an average of 11.090 years of academic age for scholars to be awarded an SBE, and that SBE scientific funding is dominated by male scholars at a high 69.00%. Considering that some scholars believe that the base of female scholars who apply is originally smaller than that of male scholars, this may not be evidence of gender discrimination in the allocation of SBE funds. On the contrary, the proportion of female SBE grants successfully funded is higher than the proportion of female scholars known to other funding committees.

| Variables | Mean | Std. dev. | Min | Max |
|---|---|---|---|---|
| Publication counts | 1.22 | 2.83 | 0.00 | 113.00 |
| Citation | 15.10 | 61.06 | 0.00 | 6120.60 |
| CiteScore | 1.15 | 2.51 | 0.00 | 49.39 |
| Year | 2009.37 | 5.36 | 1990.00 | 2019.00 |
| Gender | 0.69 | 0.46 | 0.00 | 1.00 |
| Academic age | 11.09 | 9.49 | 0.00 | 78.00 |
| ln (Grant amount) | 11.56 | 1.39 | 6.77 | 16.12 |
| Length of year funded | 3.95 | 0.99 | 3.00 | 8.00 |
| ln ( Publications $_{fields}$ ) | 11.80 | 2.70 | 1.95 | 17.09 |
| ln ( Citations $_{fields}$ ) | 14.36 | 2.57 | 3.53 | 19.60 |

[5] The U.S. News and World Report and QS lists were officially launched in 2014 and 2010, respectively, and cannot cover the time window of the SBE research funding observation period. Moreover, universities' rankings in the major lists do not generally change significantly over a few years, and this is particularly true of U.S.-based universities, whose rankings are more stable (Selten et al., 2020). We have therefore considered the rankings of these universities in the 2022 U.S. News and World Report and QS versions.

| | | | | |
|---|---|---|---|---|
| ln ( Publications$_{Affiliations}$ ) | 14.03 | 1.74 | 3.50 | 16.79 |
| ln ( Citations$_{Affiliations}$ ) | 14.03 | 1.74 | 3.50 | 16.79 |
| QS ranking | 513.54 | 621.36 | 1.00 | 2000.00 |
| USNEWS ranking | 1301.79 | 560.10 | 1.00 | 1749.00 |
| Employer reputation | 38.12 | 34.97 | 1.00 | 100.00 |

\* $Citations_t^i = \frac{\sum_{Z=0}^{z=n} Citations_{Z,t}^i}{n}$ and $Citescore_t^i = \frac{\sum_{Z=0}^{z=n} Citescore_{Z,t}^i}{n}$ denote the average citations and the average CiteScore of publication published by author $i$ in year $t$, respectively. $Age^{academic} = Year^{award} - 1 - Year^{initial}$ indicates the difference between the year before the award year and the publication year of the first WoS- or MAG-indexed publications.

**Table 1. Descriptive statistical results for variables.**

## 4. Methods

### 4.1 IV and 2SLS model

The first IV is the political resources of academia. Scholars have explored the important role that social elites play in the distribution of competitive funding. Drawing on Lawson et al. (2021) employing academic national socio-political capital as an IV for Italian scholars, we introduce it to the English-speaking countries and expect that it will affect the probability of success in accessing U.S. and international competitive funding. Particularly, in environments where resources are scarce or competitive, political hegemony will dominate the allocation of resources (Mitroff & Chubin, 1979). Thus, drawing on Lawson et al. (2021) employing academic national socio-political capital as an IV for Italian scholars, we introduce it to the English-speaking countries and expect that it will affect the probability of success in accessing U.S. and international competitive funding. Here, we redefine academic national socio-political capital for the NSF, i.e., defining the role of managers, board members, and external experts in the SBE foundation. We count the number of PIs who served in the SBE Division/NSF Board, NSF Directors, Social Sciences and Humanities Division/American Philosophical Society (APS), or American Association for the Advancement of Science (AAAS) from 1990 to 2019.[6] Although a degree of scientific achievement may be required to hold these influential roles, doing so might be largely the result of the accumulation of socio-political capital by scholars. For example, numerous U.S. presidents or U.S. Supreme Court justices, e.g., George Washington, Barack Hussein Obama, and Elena Kagan, are elected members of the APS Society.

---

[6] Appropriate IVs are required to affect scholars' performance indirectly by affecting research funding, rather than directly. As Hoenen and Kolympiris (2020) found, a history of being employed by the NSF will affect an academic's research funding in the following five years. The observed time window for scholars' performance in this study is 1995 to 2019. Therefore, we extend the scholars' hiring observation period by five years, 1990 to 2019, to determine whether the hiring profile of scholars from 1990 to 1994 may directly affect their performance.

In our dataset, 58 PIs served or are serving as members of the NSF Board of Directors or as program directors in SBE departments, or as external experts for grant review meetings or programs; 62 PIs are elected as members of the APS Association's seven fields of Sociology and Demography, Economics or Linguistics related to SBE funding areas; and 142 PIs are elected as key members of AAAS' Social and Economics Science and its interdisciplinary areas. The dummy variable measuring political capital is taken as one (1) in the first year of the lead role and in the following four years to indicate its impact on academics over time[7]. Additionally, we employed two further strategies to validate the reliability of academic political hegemony as an instrumental variable. First, we categorized scholars' political hegemony into different levels, treating these as separate treatments, and examined whether varying degrees of political capital have a statistically significant direct effect on academic performance, relative to scholars with no political capital. This approach tests whether political hegemony might directly affect academic output, potentially compromising the exogeneity assumption of the instrument. Furthermore, we hypothesized that if academic political hegemony directly impacts scholarly productivity, this effect would likely vary according to the scholar's level of political capital. To test this hypothesis, we conducted separate IV analyses for different levels of political capital and compared their impact on the estimated effect of research funding. Results show that varying levels of political capital did not lead to significant differences in the funding effect estimates, lending further support to the exogeneity assumption of academic political hegemony as an instrumental variable (see details in **"F. FURTHER ANALYSIS ON THE USE OF ACADEMIC POLITICAL HEGEMONY AS AN INSTRUMENT" of the Supplementary material**).

The second IV we selected comes from isomorphism among scholars. We draw on imitation isomorphism in organization theory (DiMaggio & Powell, 2000) and semantic isomorphism in mathematics (Moraschini, 2016) to explain the potential imitation isomorphic behavior of academics and the structural framework of isomorphism in which they are positioned.. We introduce this imitation isomorphism to individual scholars. However, within the social sciences, the scope and design of research topics by different researchers span a considerable range (Bechhofer & Paterson, 2012). We therefore further narrowed the structure of imitation isomorphism from the faculty to the research topic level. We use the topic of a scholar's funded project as the main direction of their research contribution within a given phase; the research directions of scholars under the same research topic are more similar within the social science subfield, i.e., the semantic isomorphic results of the research topic symbolization. Suppose that an individual academic belongs to an entity organization (university) $U$ and his/her funded research topic is $T$. $U = [U_1, U_2, \ldots U_m]$ indicates that there are $m$ universities after duplication of all scholars' affiliations; $T = [T_1, T_2, \ldots T_n]$ indicates the research topic to which the scholar's research grant belongs. We then count the number of award-winning scholars for each research topic for each university in the $m \times n$ matrix.

---

[7] For the robustness of the IVs, we also looked at the validity of the IVs when setting the time window to three years and seven years (see details in **"G. SHORT- AND LONG-TERM VALIDATION OF IVS" of the Supplementary material**).

When scholars within a cell, the same organization and research topic, receive research funding, other scholars within the same cell will imitate the successfully funded scholars when faced with research financial issues, research blocking, or promotion difficulties. This imitation may affect the probability of scholars who imitate others receiving grants, relative to scholars in other cells where there are no scholars to be imitated. The further discussion on the exclusion restriction and design of the imitation isomorphism instrument are shown in **"H. FURTHER DISCUSSION ON THE EXCLUSION RESTRICTION AND DESIGN OF THE IMITATION ISOMORPHISM INSTRUMENT" of the Supplementary material**.

A third IV is also chosen, namely PIs' familiarity with the SBE program and application. The NSF directorates' official NSF days and workshop events are selected as a key channel to affect scholars' familiarity with the grant application. NSF days and workshops are held individually or in collaboration with a university selected by different NSF directorates. They are designed to increase the competitiveness of scholars and universities in competing for NSF funding, highlighting and guiding proposal writing, performance review processes and priorities, and include answering questions from academics. The SBE has conducted NSF days and workshops at 98 universities, and it does not tend to select universities that have historically received more funding to host events. The variable approximating researchers' knowledge of and training in SBE is the number of events such as NSF days hosted by their university in the five years prior to their receiving the SBE grant.

We include the above three IVs in the 2SLS model:

$$F_{it} = \alpha_0 + \alpha_1 I_{it} + \alpha_2 X_{it} + \mu_{it} \qquad (1)$$

$$P_{it} = \beta_0 + \beta_1 F_{it} + \beta_2 X_{it} + \varepsilon_{it} \qquad (2)$$

In the first-stage regression (Eq. 1) aiming to estimate the probability of scholar $i$ receiving a grant in time $t$ ( $t = 0$ : scholar $i$ does not receive funding, $t = 1$ : scholar $i$ is supported by funding at $t$ and in the following four years) given the inclusion of instrumental and covariate variables, $I_{it} = \left[ Employment_{it}, Dominance_{it}, Familiarity_{it} \right]$ is a vector of IVs. $Employment_{it}$ denotes the dummy variable for scholar $i$ being employed by the NSF and associations in the social sciences at time $t$ ; $Employment_{it} = 1$ indicates the long-term effect in the year the scholar is employed and the four years thereafter, otherwise $Employment_{it} = 0$ . $Dominance_{it}$ represents the normalized inverse of the number of historical grants in the research topic at the university where $i$ is affiliated; $Familiarity_{it}$ identifies the interaction term between the number of training events such as NSF days and workshops hosted by $i$ 's university in the three years prior to the grant and the time dummy variable for those three years. $\alpha_0$ is the constant term. $X_{it}$ is a matrix of control variables for the individual, university, and research area of $i$ , including year, gender of the academic, academic age, logarithm of grant amount, logarithm of the average publication count and citation count per year for the research area to which the publication belongs, logarithm of the average publication count and citation count per year for the affiliation, QS and US NEWS ranking, and employer reputation score of the affiliation (see Table 1).The second stage (Eq.2) estimates the effect of predicted availability of SBE funding $F_{it}$ derived in the first stage from the scholar performance $P_{it}$ of scholar $i$ in time $t$ . Simultaneously, we control for the same set of control variables $X_{it}$ in Eq.1. Then, $\beta_1$ is the treatment effect of the science grant on scholars' performance that we are interested in. $\beta_0$ is the constant term.

## 4.2 Reverse causality

It is important to reiterate our desire to discover the impact of research funding on scholars' performance. However, historical scholars' performance may affect the availability of research grants to academics, i.e., a reverse causality accompanies the two. Reverse causality has been demonstrated to cause the independent variable to be associated with the random error term (Malter, 2014), resulting in biased effect estimates. Thus, we use the logarithmic mean of the scholars' performance of academics in the three years prior to the sample period as initial scholars' performance, as suggested by Wooldridge (2010). This initial performance represents the path dependence of performance and the permanent heterogeneity of PIs including academic motivation and cognitive ability (Fernández-Zubieta et al., 2016). We then include this control variable in the 2SLS model to complement the predicted effects.

**4.3 Robustness tests**

First and most importantly, we confirm whether there is an endogeneity issue with SBE funding allocation, which is the basis for the IVs to isolate bias (Mutl & Pfaffermayr, 2011). Then, this study tests the validity of the three IVs we selected, including tests of exogeneity of IVs (Hausman et al., 2005), tests of under- and over-identification (Ullah et al., 2021), and tests of weak IVs (Andrews et al., 2019).

While the IVs selected are reliable, we also need to determine the robustness of the research funding effect. In this study, all individuals in the treatment group are randomly divided into a pseudo-treatment group and a pseudo-control group; the control group is similarly divided into a pseudo-treatment group and a pseudo-control group. In general, we hypothesize that there should be no significant funding effects or scholars' performance differences between the respective two subgroups of the treatment and control groups. Therefore, we repeat the IVs and 2SLS strategies separately for the two subgroups in the treatment and control groups, as shown in Figure 1, to see if a research funding effect emerges.

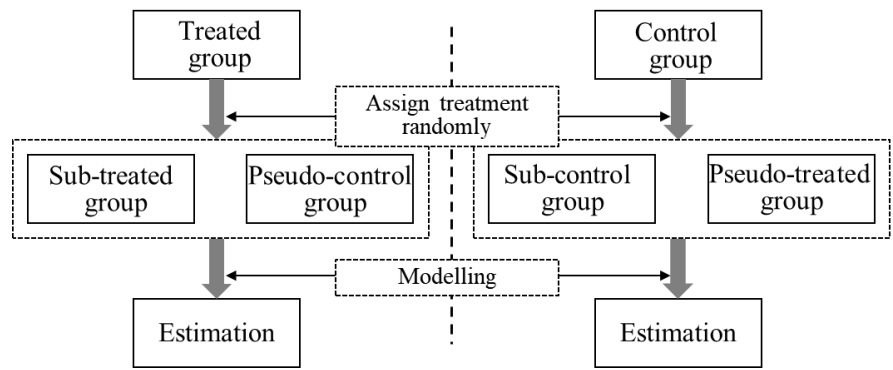

**Figure 1 Robustness tests for research funding effects.**

**5. Results**

**5.1 OLS estimation results**

Table 2 shows a considerably significant positive correlation between SBE research funding and scholars' performance, using OLS regression. Controlling for gender, affiliation, research area, and initial scholars' performance of a scholar, receiving funding resulted in a 68.9% rise in the number of journal articles authored by the scholar; a 46.669 rise in the average number of citations to a single publication; and the ability of the grantee's research to be published in higher-impact journals, where the average CiteScore rose by 5.524, compared to those scholars who had not received funding.

| Variables | Article counts | Citation counts | CiteScore |
|---|---|---|---|
| Funding | 0.689*** | 46.669*** | 5.524*** |
| | (0.043) | (1.010) | (0.336) |
| Initial ln (Article counts \| Citations \| Citescore) | 0.089*** | 22.919*** | 1.852*** |
| | (0.013) | (0.313) | (0.105) |
| Funding quota | 0.002 | 0.932*** | 0.834*** |
| | (0.015) | (0.258) | (0.083) |
| Academic age | 0.075*** | -0.165*** | 0.167*** |
| | (0.002) | (0.033) | (0.011) |
| Year | 0.086*** | -1.763*** | -0.230*** |
| | (0.005) | (0.091) | (0.030) |
| Gender | 0.243*** | -0.015 | 0.351 |
| | (0.050) | (0.834) | (0.268) |
| ln ( Publications $_{fields}$ ) | -0.165*** | -8.932*** | -5.284*** |
| | (0.023) | (0.575) | (0.192) |
| ln ( Publications $_{Affiliations}$ ) | -0.030 | -9.322*** | -6.951*** |
| | (0.042) | (0.860) | (0.281) |
| ln ( Citations $_{fields}$ ) | 0.180*** | 9.112*** | 5.533*** |
| | (0.025) | (0.605) | (0.201) |
| ln ( Citations $_{Affiliations}$ ) | -0.014 | 8.488*** | 6.655*** |
| | (0.038) | (0.757) | (0.248) |
| Employer reputation | 0.001 | 0.125*** | 0.038*** |
| | (0.001) | (0.016) | (0.005) |
| US NEWS ranking | 0.001*** | 0.001 | 0.001** |
| | (0.000) | (0.001) | (0.000) |
| QS ranking | -0.001* | 0.004*** | 0.001 |
| | (0.000) | (0.001) | (0.000) |
| Constant | -171.408*** | 3,471.758*** | 425.557*** |
| | (9.494) | (183.504) | (59.650) |

\* p < 0.05; \*\* p < 0.01; \*\*\* p < 0.001. The values in parentheses are standard errors.

**Table 2. OLS regression results.**

**5.2 2SLS estimation results**

In 2SLS first stage, as shown in Table 3, the three IVs we selected have a significant effect on the distribution of SBE grants. The coefficient of the isomorphic effect between scholars (coef.=0.004, p=0.000) indicates that, the more grants scholars in the same research topic at the same university receive, the more likely it is that subsequent scholars will receive grants. The probability of an academic receiving an SBE grant in the subsequent three years rises when they occupy an administrative position in academic associations in the social sciences and economics, with a 4.7% increase in probability. This implies that political resources and a social elite culture do exist in the allocation of scientific research resources (Hoenen & Kolympiris, 2020; Lawson et al., 2021). In addition, the greater the number of NSF days and grant application training activities conducted at universities, the greater it negatively affected the probability of the social and economic scholars at these universities receiving funding, decreasing by 0.6%. The SBE's outreach and training efforts may trap the ideas and efforts of some academics in the project application process, weakening the competitiveness of the proposal.

In contrast to OLS, the 2SLS second-stage analysis shows considerable variability in research funding's impact on scholar performance. Although the OLS underestimates the impact of scientific funding on the number of journal articles and overestimates the impact on the number of citations to publications, we observe that receiving funding does increase the number and impact of scholars' publications but does not lead to a greater possibility of publishing in higher-impact journals. We see that receiving funding resulted in 2.816 more journal articles being published by funded scholars, and the average number of citations to these publications is raised by 16.421.

| Variables | 2SLS 1st stage | 2SLS 2nd stage | | |
|---|---|---|---|---|
| | Funding | Article counts | Citation counts | Citescore |
| Funding | - | 2.816*** | 16.421** | 2.883 |
| | | (0.331) | (5.552) | (1.572) |
| Initial ln (Article counts \| Citations \| Citescore) | -0.196*** | 0.599*** | 16.885*** | 1.327*** |
| | (0.001) | (0.066) | (1.344) | (0.332) |
| Funding quota | -0.015*** | 0.001 | 0.655* | 0.804*** |
| | (0.001) | (0.012) | (0.262) | (0.080) |
| Academic age | -0.003*** | 0.075*** | -0.266 | 0.158*** |
| | (0.000) | (0.003) | (0.043) | (0.014) |
| Year | -0.013*** | 0.108*** | -2.351 | -0.269*** |
| | (0.000) | (0.006) | (0.142) | (0.040) |
| Gender | 0.039*** | 0.229*** | 1.001 | 0.449 |
| | (0.004) | (0.040) | (0.891) | (0.267) |
| ln ( Publications $_{fields}$ ) | -0.026*** | -0.381 | -10.066 | -5.354*** |
| | (0.003) | (0.026) | (1.058) | (0.202) |
| ln ( Publications $_{Affiliations}$ ) | -0.099*** | 0.298*** | -12.233 | -7.195*** |
| | (0.004) | (0.062) | (1.628) | (0.300) |
| ln ( Citations $_{fields}$ ) | 0.026*** | 0.420*** | 10.261*** | 5.603*** |
| | (0.003) | (0.026) | (1.014) | (0.211) |
| ln ( Citations $_{Affiliations}$ ) | 0.097*** | -0.291 | 11.276*** | 6.888*** |
| | (0.003) | (0.053) | (1.279) | (0.258) |
| Employer reputation | 0.000*** | 0.002** | 0.122*** | 0.037*** |
| | (0.000) | (0.001) | (0.016) | (0.005) |
| US NEWS_ranking | 0.000 | 0.001*** | 0.001 | 0.001*** |
| | (0.000) | (0.000) | (0.001) | (0.000) |
| QS ranking | 0.000*** | 0.000 | 0.003** | 0.000 |
| | (0.000) | (0.000) | (0.001) | (0.000) |
| Isomorphism | 0.004*** | | | |
| | (0.000) | | | |
| Political hegemony | 0.047** | | | |
| | (0.020) | | | |
| NSF training | -0.006*** | | | |
| | (0.001) | | | |
| Constant | 27.140*** | -218.319*** | 4,667.165*** | 506.636*** |
| | (0.782) | (12.747) | (287.981) | (81.204) |

\* p < 0.05; \*\* p < 0.01; \*\*\* p < 0.001. The values in parentheses are standard errors.

**Table 3. 2SLS regression results.**

However, research funding does not improve the likelihood of publishing in more prestigious journals. There is no statistically significant relationship (coef.=2.883, p=0.067) between scientific funding and journals' CiteScore. One key driver of this phenomenon could be the short-term performance metrics inherent in academic and funding evaluation systems. Researchers are often under pressure from funding agencies to produce results within tight timeframes and to report progress regularly, which may incentiva a focus on quantity rather than the depth or quality of research (Zacharewicz et al., 2023). This short-term effect, driven by time pressures, may lead scholars to submit their work to journals with shorter review cycles and higher acceptance rates, which are often of lower impact compared to more prestigious journals that require longer review periods and more stringent criteria (Johann et al., 2024). Moreover, the funding system itself may inadvertently encourage this behavior, as researchers, in an effort to secure continued funding, may prefer journals that allow for quicker publication, rather than engaging in the more time-consuming process of submitting to top-tier journals that require greater investment (Butler, 2003). This issue is further exacerbated by broader academic cultural dynamics. Success is often equated with quantity—such as the number of published papers—rather the quality or impact of those papers (Lawson et al., 2021). This emphasis on short-term results may undermine the cultivation of long-term research quality and innovation (Zacharewicz et al., 2023).

Building on the earlier finding that research funding does not significantly improve the likelihood of publishing in prestigious journals, further analysis reveals its heterogeneous effects across journals with different CiteScore levels. Specifically, while research funding does not significantly influence articles published in Top 10% CiteScore journals (coef.=−0.650, p=0.547), it has a substantial positive impact on publications in Bottom 10% CiteScore journals (coef.=3.183, p=0.002). This suggests that funding plays a crucial role in enhancing the visibility and impact of work published in lower-impact journals. One explanation for this pattern could be the short-term performance pressures discussed earlier, which drive scholars to prioritize rapid results and easier publication opportunities. In this context, funding may provide critical support for researchers operating in resource-constrained environments, enabling them to meet these pressures by improving the quality of their submissions to lower-impact journals. By alleviating resource limitations, funding helps raise the "floor" of academic performance, even if it does not substantially influence the "ceiling" represented by top-tier journals. This underscores the dual role of funding: while it supports research productivity broadly, its impact is most pronounced in improving the output of lower-performing publications rather than elevating work to higher-impact outlets. The regression results for Top 10% and Bottom 10% CiteScore journals and the corresponding tests for the validity of the instrumental variables used in these regressions are presented in **"I. ANALYSIS OF THE EFFECTS OF RESEARCH FUNDING ON TOP 10% AND BOTTOM 10% CITESCORE JOURNALS" of the Supplementary material**.

Regarding control variables, there are some gender and institutional gaps in scholars' performance. Receiving funding significantly increases publication counts for male academics compared to female academics, but this gender gap is to a smaller scale. There is no significant relationship between gender and scientific impact. Meanwhile, publication and impact of research areas, university rankings, and employer reputation do have significant associations with scholars' performance. For example, when winners are male, they publish 0.039 more publications compared to non-winners; however, the relationship between the gender gap and the citation count of publications and journal impact is not significant, which is consistent with the findings of other research (Benavente et al., 2012; Hottenrott & Thorwarth, 2011; Lawson et al., 2021). The higher the employer reputation score of the university, the more efficient and effective the scientific publications of academics, which further confirms the findings of a number of studies (Medoff, 2006; Way et al., 2019).

### 5.3 Validity test results for IVs

The use of IV models often requires endogeneity, exogeneity, over-identification, and under-identification tests to confirm the validity and strength of the IVs. The results are shown in row (1) of Table 4, where we first test whether SBE research funding is endogenous using the Hausman test. Therefore, we reject the null hypothesis that SBE funding is exogenous, which underlies the subsequent use of IVs to isolate endogeneity.

| | Tests | | | Article counts | Citation counts | Citescore |
|---|---|---|---|---|---|---|
| (1) | SBE funding endogeneity | Hausman test | Robust score chi2(1) | 19.811*** | 58.873*** | 9.865*** |
| | | | Robust regression $F$ | 20.022*** | 59.376*** | 9.892** |
| (2) | Over-identification | Hansen J test | Statistics | 0.611 | 4.709 | 1.222 |
| (3) | Under-identification | Kleibergen-Paap rk LM test | Chi-sq(3) | 1279.741*** | | |
| (4) | Weak identification test | Cragg-Donald Wald test | $F$ statistic | 507.350 | | |
| | | Stock-Yogo weak ID test | 5% maximal IV relative bias | 13.910 | | |
| (5) | 2SLS first stage | Robust F | | 544.236*** | | |
| | | Minimum eigenvalue statistics | | 507.350 | | |
| | | 2SLS size of nominal 5% Wald test (10% critical values) | | 22.300 | | |
| | | 2SLS size of nominal 5% Wald test (15% critical values) | | 12.830 | | |

* $p < 0.05$; ** $p < 0.01$; *** $p < 0.001$.

**Table 4. Results of validity and robustness tests of IVs.**

We then use Wooldridge's (1995) robust score test, Hansen $J$ statistics, and Kleibergen-Paap rk LM statistic to determine whether IVs isolating SBE endogeneity are over- or under-identifying, with the results shown in Table 4 rows (2) and (3). We find that both tests show that the null hypothesis that the variables are exogenous cannot be rejected. Meanwhile, the Cragg-Donald Wald test and Stock-Yogo weak ID test are used to test for weak IVs, with the null hypothesis that the IVs are weak IVs. Usually, when the F-statistic of the Cragg-Donald Wald test is greater than the 5% maximal IV relative bias of the Stock-Yogo weak ID test, then we can be confident that there are no weak

IVs. The results are shown in row (4) of Table 4, where the test significantly rejects the null hypothesis, implying that the three IVs are strongly correlated with the endogenous variables.

Further, we report the results statistics for the first stage of 2SLS. As shown in row (5) of Table 4, the Robust F-statistic and the minimum eigenvalue statistics for the first stage are 544.236 and 507.350, respectively, both of which are greater than the test values suggested by other scholars that the threshold $F$ should be greater than 10. Or, when both the robust F-statistic and the minimum characteristic statistic are greater than 10% or 15% of the critical values of the 2SLS size of nominal 5% Wald test, we can conclude that there are no weak IVs. Our robustness tests for the funding effect are presented in **"J. ROBUSTNESS TEST RESULTS FOR SBE FUNDING EFFECTS" of the Supplementary material**.

## 6. Discussion

In this study, we revisit the relationship between research funding and scholars' performance using the example of 9,501 PIs supported by the SBE, the largest social science funding agency in the U.S. Particularly, we introduce three IVs to isolate the endogeneity issues of research funding effect estimates and consider characteristics of individual researchers, affiliations, and research areas in combination with the 2SLS model to explain the effects of SBE support for the social science community. Robustness tests of both these IVs and the estimation of funding effects confirm our key results.

To adequately address the endogeneity issue that is challenging in the existing literature, we highlight the theoretical background and validation process of the three IVs, the political hegemony of academics, imitation of isomorphic behavior, and program familiarity. This is one of the essential contributions of this study. We find that scholars' political capital significantly increases the probability of scholars receiving funding, consistent with previous findings (Hoenen & Kolympiris, 2020; Lawson et al., 2021; Viner et al., 2004) that political hegemony increases scientists' competitive advantage. Imitation isomorphic behavior and conditions among scholars similarly contribute to the probability of grant receipt. On the one hand, information symmetry is critical when applying for grants, and the conditions and circumstances under which imitation occurs among scholars may contribute to the information fluidity of grant applications by imitating scholars. On the other hand, as more grant successes arise in an environment where imitation is isomorphic, i.e., where there is a cumulative advantage in research grants for particular research topics at the given university, research grants are more likely to be allocated to those universities with a history of good application performance. This supports the findings of several studies (e.g., Perc, 2014; Steinþórsdóttir, Einarsdóttir, Pétursdóttir, & Himmelweit, 2020) demonstrating a Matthew effect in applying for and receiving research grants.

Yet, worryingly, the SBE's project writing training and outreach activities at universities appear to have a negative effect on the success rate of grant applications. When SBE institutions hold more of these events at universities, it rather inhibits the possibility of scholars from these universities receiving funding. The fact that funding agencies tend to choose universities with a weak funding history to host such events may further inhibit the ideas of academics at these universities and even mislead them to focus more on the normative aspects of the proposal than on other key factors. Previous research has demonstrated that 19 or more factors have been found to contribute to the success of applications, including teamwork, clear goals, and communication skills (Beleiu et al., 2015). Therefore, researchers from these universities require more holistic support than just training in writing and receiving publicity, as if they need five fingers working together to make a fist for maximum impact, rather than being taught to focus their strength on just one finger.

After using these three IVs and considering reverse causality, our study shows that funding can affect scholar performance, supporting the results from Ebadi and Schiffauerova (2016) and Pagel et al. (2015). However, research funding may have a limited effect—e.g., it does not enable recipients' knowledge to be published in more prestigious journals. This might be attributable to the system of research evaluation and short-termism. Scholars receiving research grants are frequently required to report or evaluate their performance to the funding agency on an annual or regular basis, and are even required to produce results within a short period of time. This "distorted" system of evaluation and requirements for funding or grants has been shown to cause considerable pressure on researchers, who complain about the insufficient time to publish in more challenging journals (Groen-Xu et al., 2021). Thus, by abandoning highly prestigious journals that tend to have longer review cycles, are more difficult to publish in, and require more of a researcher's time commitment (Paiva et al., 2017), academics may be forced to present their knowledge in journals that are less prestigious or more familiar. More worryingly, O'Regan and Gray (2018) confirmed that the review system and short-termism of performance evaluation systems in funding programs convey this short-termism to academics, resulting in less reproducible research and less exploration of more unknown knowledge. We would thus appeal to governments and funding agencies to continue optimizing their funding evaluation systems by proposing more realistic and feasible funding and evaluation strategies with a stronger focus on the long-termism required for knowledge quality.

We need to acknowledge some limitations. Our study focuses on regular grants, in order to avoid the interference of other research grant types such as student scholarships, infrastructure purchases, and non-competitive grants in the evaluation. Therefore, our findings may not be applicable to irregular types of research grants. This study uses the United States, one of the English-speaking countries, as an example, and it remains to be determined whether our IVs and findings are applicable to other English-speaking countries that, in particular, have different funding and award structures than the United States. In addition, our study observes changes in the overall performance of scholars in the five years prior to and five years following the award, making it challenging to capture the potential impact of research grants on scholars' further academic careers and the annual dynamics of grant effects.

## 7. Author contribution

Y.D.: Conceptualization, data curation, formal analysis, funding acquisition, instigation, methodology, visualization, writing – original draft.

Y.B.: Conceptualization, funding acquisition, investigation, methodology, project administration, resources, supervision, writing – review and editing.

## 8. Competing interests

The authors declare no competing interests.

## 9. Funding information

Yi Bu is supported by the National Natural Science Foundation of China (#72474009, #72104007, and #72174016). Yang Ding is supported by the Ph.D. fellowship (No. s2222886) in Management Science and Business Economics from the University of Edinburgh Business School.

## 10. Acknowledgments

An early version of this paper was presented at Wuhan University and Sun Yat-sen University in 2023 and 2024, respectively; the authors thank the suggestions from the audiences. The authors are also quite grateful to the editor and the anonymous reviewers for their constructive comments.

## 11. Data Availability

NSF SBE grant data: NSF funding data for the Social, Behavioral, and Economic Sciences (SBE) Division is publicly available at https://www.nsf.gov/funding/opportunities.

Data on scholars' political hegemony: Information about scholars holding significant positions, such as NSF Board members and SBE division program directors, remains publicly accessible through sources like https://www.nsf.gov/nsb and https://www.nsf.gov/sbe/sbe-advisory-committee#members-ba9. However, information about external experts or scholars involved in major SBE budget decisions and related matters is now restricted (e.g., https://new.nsf.gov/events/iucrc-center-e-design-edesign-iab-meeting-5/2019-11-05). Due to significant updates to NSF's website and policies in 2022, access to information about earlier events, particularly those before 2019, has become substantially restricted. Since 2019, detailed records on event participants, including internal and external advisors and reviewing scholars, are no longer available for public viewing. This policy change is likely intended to protect sensitive information, including proprietary technical data, financial details, and personal data associated with grant proposals, in compliance with the Government in the Sunshine Act (5 U.S.C. 552b(c), Sections 4 and 6). As an alternative, NSF now provides video records of key meetings, which remain accessible to the public. Scholars interested in NSF events may gather legally available information from these recordings.

Data on membership in relevant fields of the American Philosophical Society (APS) such as Sociology, Demography, Economics, and Linguistics can be accessed via https://search.amphilsoc.org/memhist/search. Data on the American Association for the Advancement of Science (AAAS) members in the fields of Social and Economic Sciences and interdisciplinary areas can be accessed through https://www.aaas.org/membership.

Data on NSF Days: Information about NSF Days can still be accessed through https://www.nsf.gov/od/olpa/eventgroups/nsf-days. However, starting from 2022, due to the relevant exemptions under the Sunshine Act, NSF has hidden the historical records of NSF Days. Nonetheless, the historical information about NSF Days and similar events can still be retrieved through the official websites of various universities.

Bibliographic data sources: Bibliographic data were obtained from Microsoft Academic Graph and Web of Science. Microsoft Academic Graph data is accessible at https://www.microsoft.com/en-us/research/project/open-academic-graph/, though updates ceased following the sale of the platform Web of Science data is available via institutional subscriptions. We provide detailed guidance on joint searching of Web of Science and Microsoft Academic Graph bibliographic databases to obtain a more comprehensive publication dataset for scholars (see " E. JOINT SEARCHING OF WEB OF SCIENCE (WOS) AND MICROSOFT ACADEMIC GRAPH (MAG) BIBLIOGRAPHIC DATABASES" of the Supplementary material).

## 12. Supplementary materials

Please see Supplementary Materials for Section A to J in this study.

# SUPPLEMENTARY MATERIALS FOR
# POLITICAL HEGEMONY, IMITATION ISOMORPHISM, AND PROJECT FAMILIARITY: INSTRUMENTAL VARIABLES TO UNDERSTAND FUNDING IMPACT ON SCHOLAR PERFORMANCE

## A. MOTIVATION

### A1 Alternative metrics of scientific productivity

The development of information retrieval and science citation indexing has made the publication records and citation counts of academics visually accessible, resulting in the possibility of quantifying the scholars' performance (Sanderson & Zobel, 2005). After World War II, the seemingly uncontrolled expansion of knowledge in science began. The following decades witnessed a revolution in the medium of knowledge from physical papers to electronic completion. For example, in 1997, in less than a year, NASA's digital library publications material read exceeded the number of paper publications read by all astronomy libraries combined in history (Kurtz & Bollen, 2011). As per our search on WoS, by 2021, the number of countries in the world with more than one million electronic publications included in the WoS core collection rose to 11; the number of publications included in the WoS in the 20 years since the start of the 21$^{st}$ century is more than 7.169 times the number of electronic publications in the entire 19$^{th}$ century.

On the other hand, academics recognize that scholars' performance should not only be about the quantity of research, but also the quality of research. Citation counts, for example, are becoming increasingly available as an evaluator of articles, individuals, journals, research institutions, and even entire countries (Ding et al., 2021; Jacob & Lefgren, 2011), complementing the potential one-sidedness of publication counts for performance explanations. Also, the citation impact of publication such as CiteScore and Journal Impact Factor are also beginning to be used as alternative measures of the research impact of scholars and the prestige of journals (McKiernan et al., 2019).

### A2 Research funding effects exploration process

The study of the relationship between research funding and scholars' performance in the 19$^{th}$ century is referred to by some scholars as a "stimulus-response" model (Kevles, 1977). In other words, research funding is defined as a government decision that permeates governments, universities, and research institutions, as well as individuals. The evaluation of the outcomes of "mediocre" and "good" science after funding influences government decisions and subsequent funding, potentially resulting in the elimination of "mediocre" scientists, particularly in times of fiscal constraints (Rothenberg, 2010). However, the prioritization of scientific funding in various research fields is frequently influenced by the efficiency with which disciplinary knowledge is translated into performance, as well as by politicians. Academics are also becoming aware that the arbitrarily slashing of so-called "mediocre" areas of science in an effort to save money may have the opposite effect on scholars' performance (Mutz et al., 2015), particularly in an era of growing interdisciplinarity. In the meantime, governments, recognizing the significance of each discipline, is increasing its investment in research, though not at the same rate in all fields (House of Commons of the United Kingdom, 2019). Nonetheless, as the influx of researchers exceeds the increase in funding, areas with lower funding increases may amount to a "cut" in research funding. To rationalize the increase in funding and to mitigate the policy bias that restricts the development of certain research areas, it is crucial to evaluate the performance of funded science.

A number of studies have suggested ways to investigate the relationship between research funding and scientific output. Researchers have developed "cross-sectional'" and "longitudinal" evaluation mechanisms at the publication, individual, university, and national levels for assessing scholars' performance based on research funding (Fursov et al., 2016). The cross-sectional assessment of research funding requires researchers to compare the "mean" difference between these two indicators for two groups of academics: those who are supported by research funding and those who are not (Alkhawtani et al., 2020). UK Research and Innovation (UKRI), the largest science foundation in the U.K., states almost annually in its annual report that the publications produced by the programs it funds have a greater number of citations than those of other scientists worldwide. On the other hand, typically, longitudinal evaluations are used to measure the increase in performance of academics who have received award funding. Recent research has begun to combine these longitudinal and cross-sectional methods, comparing changes in the past and future accomplishments of academics who are successful in receiving investment to those who are not (Hanna Hottenrott & Lawson, 2017). This enables researchers to measure differences in scholars' performance between scholars or articles with distinct characteristics by incorporating additional controllable variables; for example, a chain analysis strategy using pre-award-funding-application-post-award of scientific funding and taking gender disparities and childcare responsibilities into consideration (Lawson et al., 2021).

**A3 Paradoxes in the assessment of research funding effects**

Despite the fact that some studies provided evidence that it is worthwhile exploring the relationship between research funding and scientific performance, the research funding effect still appears to be controversial. We list in Table S1 some studies and their results that explore the impact of research funding on scholars' performance. Perhaps the biggest controversy is whether research funding stimulates scholars' performance. Numerous studies conclude that government investment affects both the quantity and quality of scholars' performance. Pagel and Hudetz (2015) found a significant positive research funding effect on the numbers of scientific publications and citations, using the 397 Anaesthesiology scientists who received FAER/NIH funding from December 2014 to January 2015 as an example. Similar evidence was found in research funding in Canada, the U.K., Norway, and other nations (Ebadi & Schiffauerova, 2016). However, some academics have also expressed doubts about the research funding effects. For example, using regression discontinuity, Benavente et al. (2012) investigated over 3,000 academics and found that receiving funding from the Chilean National Science and Technology Research Fund (FONDECYT) did not affect the research quality of recipients from 1988 to 1997.

| Funding | Data | Fields | Method | Metrics | Results | Authors |
|---|---|---|---|---|---|---|
| **FONDECYT** | 3,142 scholars | Any discipline | Regression discontinuity | Publications | Positive | Benavente et al. (2012) |
| | | | | Citations | Not significant | |
| **FONCYT** | 323 projects | Any discipline | Difference in difference | Publications | Positive | López, Ubfal, Chudnovsky, and Rossi (2008) |
| | | | | Journal impact factor | Positive | |
| **German industry funding** | 678 professors | Engineering disciplines | Poisson regression | Publications | Negative | Hottenrott and Thorwarth (2011) |
| | | | | Citations | Negative | |
| **UK public funding** | 807 UK academics | Biological, mechanical and electrical disciplines | Two-way fixed effects | Publications | Positive | (Hottenrott and Lawson (2017) |
| | | | | Citations | Positive | |
| **UK industry funding** | | | | Publications | Negative | |
| | | | | Citations | Negative | |
| **Canada funding** | 3,312 scientists | Health, Natural Sciences and Engineering | OLS | Citations | Positive | Beaudry and Larivière (2016) |
| **Any competitive fund** | 276 academics | Physics and Chemistry | Instrumental variables | Publications | Not significant | Lawson et al. (2021) |
| | | | | Citations | Not significant | |
| | | | | CiteScore | Not significant | |
| **Canadian NSERC fund** | 36,124 academics | Natural Sciences and Engineering | Negative binomial regression | Publications | Positive | Ebadi and Schiffauerova (2016) |
| | | | | Citations | Positive | |
| **International funding** | 35 OECD economies | Any discipline | Negative binomial regression | Citations | Positive | Leydesdorff, Bornmann and Wagner (2019) |
| | | | | Citations | Negative | |
| **Government funds** | 35 OECD economies | | | Citations | Positive | Wagner, Travis, Jeroen, and Koen (2018) |
| **NIH** | NIH funds from 1980 to 2009 | Medical-related topics | Statistical description | Publications | Positive | Boyack and Jordan (2011) |
| | | | | Citations | Positive | |
| **NIH** | 365,380 grants | Biomedicine | Statistical descriptions | Citations | Positive | Li et al. (2017) |
| **NIH** | 600 publications | Radiology | Mann-Whitney $U$ test | Citations | Not significant | Alkhawtani et al. (2020) |
| **NIH R01 grants** | 1,184 proposals | Medical-related topics | T-test | Publications | Not significant | Wang et al. (2019) |
| | | | Matching Regression discontinuity | Citations | Negative | |
| **NIH Postdoctoral Training Grant** | 13,426 postdocs | Biological sciences and social sciences | Regression discontinuity | Publications | Positive | Jacob and Lefgren (2011) |
| **FAER/NIH funding** | 397 academics | Anaesthesiology | Statistical descriptions | Publications | Positive Positive | Paul et al. (2015) |

**Table S1 Some available studies and their controversial results.**

Among complicated and interconnected reasons for this paradox would be the different purposes of science funds. In other words, science funds are established to encompass a variety of purposes, including regular research, philanthropy, student training, infrastructure and equipment procurement, and non-competitive research projects. Some studies appear to conflate different types or purposes of investment as homogeneous research grants or competitive grants (Oliveira et al., 2019). For example, Li and Sampat (2017) mix all types of National Institutes of Health (NIH) research grants together for evaluation and claim that the grants are effective in promoting the scientific output and citations. Yet, when Wang et al. (2019) focused their study on 1,184 proposals from the NIH R01 program, they found that NIH grants failed to improve publication counts of awardees and instead stimulated high-impact research of non-awardees. This makes it possible to compare *apples* and *oranges* when assessing conflated multi-natured funds (Kaiser, 2019), resulting in a biased estimate of funding effects.

Inappropriate evaluation strategies may also contribute to the disparate outcomes of scientific funding. Researchers include non-funded academics in order to compare them to funded academics, resulting in problems such as possible selection biases. On the one hand, a sizeable proportion of academics do not wish to apply for grants due to teaching, parenting, and other family responsibilities, i.e., they oppose scientifically funded interventions. It is challenging to screen these academics (Carvan, 2022). Comparisons between these academics and award recipients may have caused severe endogenous issues. Scholars who are willing but unsuccessful in their applications may themselves have significant heterogeneity such as academic ability with those who receive funding, exacerbating the incomparability between the two (Van den Berg et al., 2009). Also, it appears that research investment is flowing to specific research topics, individuals, universities, and research institutions in order to satisfy governments (Shin et al., 2022). Johns

Hopkins University scholars have received the most research funding in the U.S. for consecutive years, while community college scholars have struggled to obtain funding.[8] Also, many foundation committees often aim to fund multiple subject areas. The wide variation in publication efficiency, impact, and stage of development between research areas may result in individual publication and impact of funded scholars in diverse research areas being at odds (Vaughan & Shaw, 2005). The establishment of these comparisons appears to disregard the scholars' performance gap between affiliated institutions and the fields of study themselves, thereby creating a potential assessment strategy failure.

## B. SOCIAL, BEHAVIOURAL AND ECONOMIC RESEARCH FUNDING (SBE)

The SBE, one of the directorates of the NSF, first funded a research project in the social sciences on 18 September 1978. Richard Levin, an economist from Yale University, was awarded $54,953 to support his work on technology and dynamic competition in market structures. In the decades since, the SBE has supported the basic research of around 5,000 academics and students each year to generate fundamental understanding of diverse research topics and stimulate new ideas to advance knowledge discovery and dissemination.[9] From providing solutions to unemployment, protecting the well-being of children, analyzing climate change, to averting terrorist aggression, the SBE directorate has funded 10 major areas of research, including anthropology, economics, geography, psychology, and sociology, to help understand the causes and consequences of behavior in human societies.

Although the SBE directorate accounts for a fairly limited annual budget for the NSF Board, it is the primary vehicle for federal research funding expenditures in the social sciences and economics.[10] In FY 2021, for example, the SBE actually received approximately $280 million in federal funding relative to the other eight research directorates, representing only about 3% of all NSF grants. Yet, the SBE received 7,146 applications for programs worldwide, among which 1,556 research projects were ultimately funded, representing a funding rate of approximately 21.8% (SBE, 2021). Also, the SBE annually publishes in advance the major academic frontiers and investment directions for

---

[8] https://ncsesdata.nsf.gov/profiles/site?method=rankingbysource&ds=herd

[9] https://www.nsf.gov/sbe/about.jsp

[10] https://www.nsf.gov/about/

the following fiscal year to support basic research that advances key national priorities. Between FY 2021 and FY 2023, the SBE has significantly increased funding for basic research in the social sciences related to climate change, artificial intelligence, and American infrastructure, and decreased funding for research in the social sciences related to biotechnology. Considering the research potential of U.S. early-career scholars, the SBE also devotes itself to increasing funding for postdoctoral researchers.

Some evidence suggests that social acceptance of government-supported NSF basic research is significant (National Science Board, 2018), but it may take years for the end use and effects of funding to become publicly known. The evaluation of SBE funding comes from the NSF's annual official report, which includes information on the chronology of SBE-funded social science fields, progress in optimizing program mission requirements, and disclosure of SBE-produced Nobel laureates in economics. The organization and implementation of regulations and compliance mechanisms that enhance the performance of researchers is also a major argument in NSF's testimony at the Congressional hearings. Nonetheless, there is relatively little evidence from academia on the effectiveness of SBE directorate funding. Understanding the profile of SBE investment and feedback to U.S. social science and economics therefore appears urgent.

## C. GENDER IDENTIFICATION BASED ON THE NAME OF THE PRINCIPAL INVESTIGATORS

We use Gender-API.com, the largest platform on the internet, for identifying gender by name and full name, to identify the gender of all academics. Gender-API covers a global database of over six million names and their gender. By invoking the *GenderApiClient* package[11], we are allowed to determine the gender of the scholar by first name, full name, and nationality. The return value delivers the gender identification result, the probability, and the number of different gender counterparts in the database.

The results of the gender identification of scholars relied on in this study are shown in Figure S1(a). Further, we show the results of the distribution of scholars with different identification accuracies (see Figure S1(b)), preserving the scholars with 95% gender accuracy calculated in the Gender-API database. A sample of 378 scholars with accuracy below the threshold is excluded.

---

[11] https://gender-api.com/en/clients

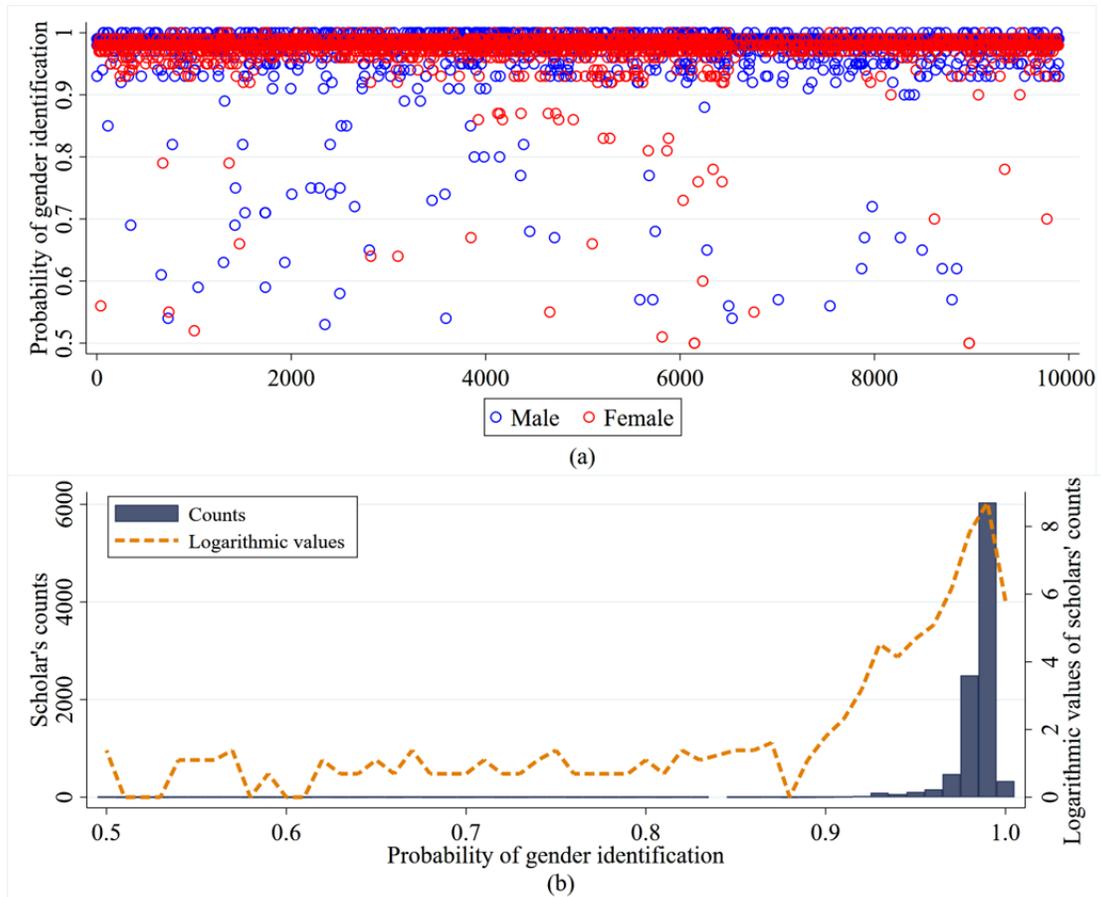

(a)

(b)

**Figure S1 Gender identification results.** The top subfigure represents the gender results for each PI and their probabilities; the bottom subfigure shows the probability distribution of the gender identification results for all PIs.

## D. TOPIC EXTRACTION OF FUNDED PROGRAMS THROUGH THE LDA MODEL

The LDA topic model was first proposed by Blei, Ng and Jordan (2003) and consists of a three-layer structure of lexical items, topics, and documents. Its basic idea is to consider a document as a mixture of various implied topics, each topic being a probability distribution of some lexical items associated with that topic, i.e., a topic-lexical item probability distribution.

In this study, the titles and program descriptions of the projects funded by the social science funds were selected as the corpus of LDA, denoted as $S$ [12]. Then the title and program description of each funded project can be considered

---

as one document in the corpus $S$. We assume that $N$ represents the length of the document of the funded project and the document length as a Poisson distribution (i.e., $N \sim Poisson(\xi)$; $\xi$ is a positive real number and is set to the same length as the "typical" document. Suppose that the vector $M$ represents the topic probability distribution of the funded documents and $M \sim Dir(v)$. $v$ is a *Dirichlet* distribution parameter. There are topics $z_n$ and $z_n \sim Multinomial(\boldsymbol{\eta})$. Then there is that,

$$p(w_n / z_n, \boldsymbol{g}) \tag{3}$$

where $w_n$ represents the lexical items that make up the funding topic $z_n$. $\boldsymbol{g}$ is a $K \cdot V$ matrix; $K$ denotes the number of topics; $V$ indicates the number of lexical items that compose a topic. Then the probability that a particular funding topic $z_i$ generates a particular lexical item $w_j$ is that,

$$g_{ij} = p(w_j = 1 / z_i = 1) \tag{4}$$

where $\eta$ is a $K$-dimensional Dirichlet random variable whose probability density equation is that,

$$P(\eta / v) = \frac{\Gamma(\sum_{i=1}^{k} v_i)}{\prod_{i=1}^{k} \Gamma(v_i)} \eta_1^{v_1 - 1} \cdots \eta_k^{v_k - 1} \tag{5}$$

The formula for the joint distribution with $\boldsymbol{\eta}$ funding topic $z_n$ and lexical item $w_n$ is as follows,

$$P(\boldsymbol{\eta}, \boldsymbol{z}, \boldsymbol{w} / \boldsymbol{v}, \boldsymbol{g}) = p(\boldsymbol{\eta} / \boldsymbol{v}) \prod_{n=1}^{N} p(z_n / \boldsymbol{\eta}) p(w_n / z_n, \boldsymbol{g}) \tag{6}$$

then, integrating over $\boldsymbol{\eta}$ and summing over $\boldsymbol{z}$, we can obtain the probability of all funded documents by summing over marginal probabilities, calculated as follows,

$$P(S / \boldsymbol{v}, \boldsymbol{g}) = \prod_{s=1}^{M} \int p(\boldsymbol{\eta}_s / \boldsymbol{v}) (\prod_{n=1}^{N_m} \sum_{z_{sn}} p(z_{sn} / \boldsymbol{\eta}_s) p(w_{dn} | z_{dn}, \boldsymbol{g})) s\boldsymbol{\eta}_s \tag{7}$$

Finally, we categorize the 9,501 funded programs into 30 funding topics, each consisting of five keywords.

___________________

Furthermore, we count the number of projects included in each topic and the distribution of gender-specific academics within these topics. The categorization and statistical results are shown in Table S2.

| Topic | Keywords | | | | Counts | Percentage |
|---|---|---|---|---|---|---|
| 0 | religious | history | immigration | aids | region | 22 | 0.24% |
| 1 | women | gender | status | race | labor | 137 | 1.46% |
| 2 | american | ethnic | migration | african | identity | 35 | 0.37% |
| 3 | human | species | genetic | evolution | data | 407 | 4.35% |
| 4 | health | medical | care | china | disease | 30 | 0.32% |
| 5 | legal | law | rights | violence | court | 151 | 1.61% |
| 6 | scholars | conference | international | scientists | issues | 352 | 3.76% |
| 7 | program | training | graduate | undergraduate | education | 211 | 2.25% |
| 8 | children | development | family | parents | adults | 174 | 1.86% |
| 9 | policy | economic | countries | international | prediction | 430 | 4.59% |
| 10 | intellectual | impacts | merit | broader | support | 32 | 0.34% |
| 11 | data | survey | social | analysis | time | 534 | 5.70% |
| 12 | risk | effects | income | health | household | 222 | 2.37% |
| 13 | environmental | change | land | climate | water | 832 | 8.89% |
| 14 | spatial | objects | infants | action | categories | 17 | 0.18% |
| 15 | market | firms | labor | economic | financial | 287 | 3.07% |
| 16 | history | century | modern | scientists | technology | 186 | 1.99% |
| 17 | people | experiments | decision | test | game | 372 | 3.97% |
| 18 | language | linguistic | speech | speakers | english | 630 | 6.73% |
| 19 | learning | cognitive | memory | human | knowledge | 320 | 3.42% |
| 20 | social | behavior | conflict | individuals | theory | 328 | 3.50% |
| 21 | technology | policy | organizations | knowledge | innovation | 863 | 9.22% |
| 22 | public | media | attitudes | survey | policy | 121 | 1.29% |
| 23 | archaeological | site | data | region | support | 447 | 4.78% |
| 24 | brain | visual | neural | human | perception | 662 | 7.07% |
| 25 | political | party | democratic | politics | parties | 93 | 0.99% |
| 26 | altitude | pathology | health | adapt | risk | 21 | 0.22% |
| 27 | models | theory | methodology | analysis | develop | 807 | 8.62% |
| 28 | social | local | cultural | community | cross | 767 | 8.19% |
| 29 | sign | deaf | varieties | heat | asl | 11 | 0.12% |

**Table S2 Results of topic extraction for SBE-funded projects and their frequency distribution.** All letters have been lowercased.

## E. JOINT SEARCHING OF WEB OF SCIENCE (WOS) AND MICROSOFT ACADEMIC GRAPH (MAG) BIBLIOGRAPHIC DATABASES

The WoS database includes details of the publication's funding agency and grant no., allowing the SBE database and WoS to be linked more accurately, as shown in Step 1 in Figure S2.

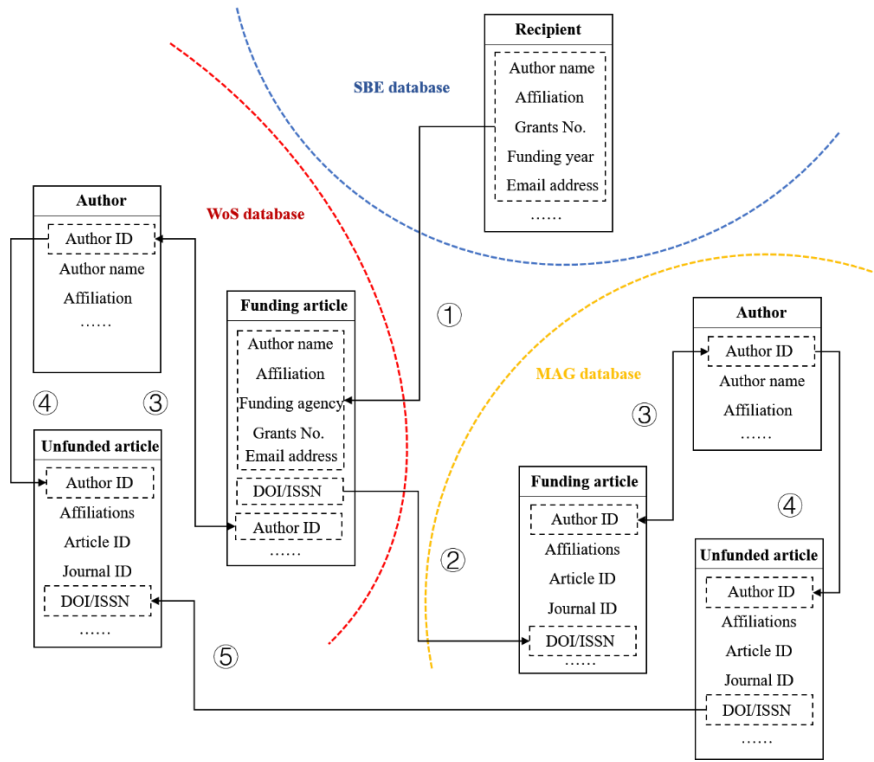

**Figure S2 A joint search process for grants, WoS and MAG databases.**

MAG does not include funding information for publications. However, it covers over 225 million publications through dynamic graphs with a rapidly expanding topology (K. Wang et al., 2020), avoiding publication statistics bias caused by different editions of the same article. In addition, MAG is proven to provide higher coverage of citations than Scopus or WoS for similar data in fields such as engineering, social sciences, and humanities (Martín-Martín et al., 2021). We only need to draw on a particular article by a given recipient that is included in both the WoS and MAG databases jointly (see Step 2 in Figure S2). The recipient's publication and citation data in both databases is then retrieved by the recipient's ID number in the respective database for this article (see Step 3 and 4 in Figure S2).

We compare the titles of publications in the WoS and MAG databases and take their concatenation to satisfy the maximum approximation to the real output of scholars. Also, we fetch the maximum number of citations for the same article in two databases.

## F. FURTHER ANALYSIS ON THE USE OF ACADEMIC POLITICAL HEGEMONY AS AN INSTRUMENT

This appendix provides additional robustness analysis for using academic political hegemony as an instrumental variable. Specifically, we examine whether academic political hegemony might directly affect academic performance, impacting its exogeneity as an instrument. Our analysis follows a structured approach to address this consideration. First, we categorize scholars' academic political hegemony into high and low tiers, represented by binary variables where 0 denotes no political resources, and 1 indicates the presence of political resources. Scholars with high or low political resources were individually assigned a value of 1 for regression purposes, enabling separate assessments. This setup allows us to understand the effect of differing levels of academic political hegemony on scholarly performance.

Our hypothesis is based on classic theoretical perspectives that suggest political capital could affect academic outcomes. Richardson (1986) social capital theory posits that individuals in influential positions accumulate symbolic capital, which grants them unique access to resources and networks. For scholars, this implies that those occupying prominent roles in academic organizations, such as NSF, APS, or AAAS, might benefit from enhanced productivity and visibility due to their positions. Higher-status bureaucratic roles provide access to resources and decision-making authority, leading to advantages in fields like academia, where higher roles might facilitate access to funding, collaboration, or high-impact publications (Weintraub et al., 1948). From these theoretical insights, we posit the following:

**Hypothesis 1:** If academic political hegemony exerts a direct influence on scholarly productivity, scholars should exhibit a more pronounced increase in academic output upon acquiring a high level of political capital compared to those attaining only a low level.

### F1 Classification of academic political hegemony

In this study, we classified academic political hegemony into "high" and "low" levels based on scholars' roles within three major academic organizations. Scholars with "high" political hegemony (*Politician 1*) are those in influential roles such as board members in NSF, APS, or AAAS. These roles involve overseeing organizational strategy, project approval, policy development, and resource allocation. Thus, scholars in these roles are considered to have substantial political capital due to their resource access and networks.

Conversely, "low" political hegemony (*Politician 2*) includes scholars with general membership or basic positions in these organizations [13]. Although these roles allow for participation, their influence over decision-making remains limited. For instance, general reviewers at NSF participate in evaluating proposals but do not typically engage in strategic decision-making or resource allocation. Similarly, members or Fellows of APS and AAAS are primarily involved as standard members without governance roles. Consequently, these scholars possess limited political capital, representing the "low" tier.

To test this hypothesis, we employed the TWFE model to evaluate the impact of academic political hegemony on three scholarly performance outcomes. This model includes controls similar to those in the 2SLS estimation, such as initial productivity, gender, academic age, funding amount, institutional reputation, and field-specific variables.

## F2 Results

The regression results reveal a consistent finding: whether academic political capital is held in higher (Politician 1) or lower (Politician 2) positions, it has no statistically significant impact on academic performance as shown in Figure S3. Across all models, the estimated coefficients for both political positions are close to zero, with confidence intervals that include zero, indicating no significant statistical relationship between political capital and academic performance. This non-significant difference applies to all academic performance indicators, suggesting that academic political capital does not confer a notable advantage in research output, citation impact, or journal quality. Instead, other factors such as initial productivity, funding levels, institutional reputation, and field-specific backgrounds play a more significant role in academic success. For instance, initial productivity stands out as a strong predictor of future academic performance, indicating that scholars with early research achievements tend to maintain their momentum, independent of political positions, continuing to publish prolifically and garner citations.

---

[13] Data on scholars' participation in NSF project reviews and discussions, for instance, was once accessible through NSF's publicly available events documentation, such as at https://new.nsf.gov/events/iucrc-center-e-design-edesign-iab-meeting-5/2019-11-05. However, significant updates to the NSF website and its policies in 2022 have made access to information on earlier events, especially those before 2019, substantially restricted. Since 2019, key meeting records are no longer available for public viewing, meaning that detailed information about meeting participants, both internal and external advisors, and reviewing scholars is currently inaccessible. This is likely because NSF's proposal review panel meetings are now closed to the public to protect sensitive information. These meetings may involve proprietary or confidential information, such as technical data, financial details like salaries, and personal data related to individuals associated with the proposals. According to the "Government in the Sunshine Act" (5 U.S.C. 552b(c), sections 4 and 6), such information is exempt from public disclosure (source: https://new.nsf.gov/events/proposal-review-panels). As an alternative, we suggest a new yet time-intensive approach: NSF has shifted documentation of key meetings to video formats, which the public can access. Through these recorded videos, it remains possible to gather legally available information on scholars' participation in NSF events.

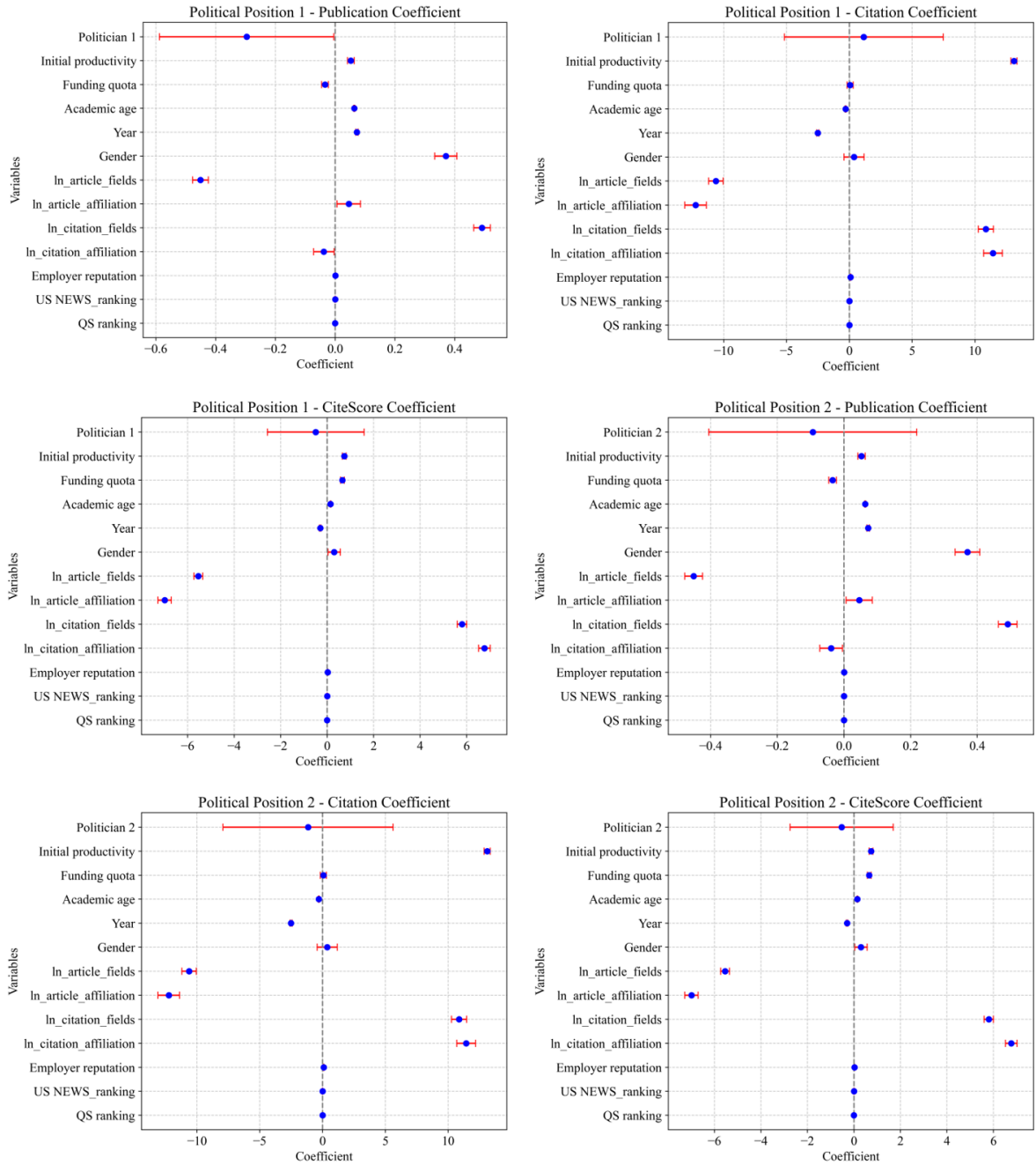

**Figure S3 Regression results for high and low political capital groups across three academic performance metrics.**

Although theories in sociology and political science posit that those in power positions might leverage their impact to increase academic output, our correlation analysis suggests that this symbolic capital does not translate into substantial academic performance advantages.

Further analysis assessed the impact of academic political capital at different levels when used as an instrumental variable for academic performance. Results indicate minimal significant differences between instrument variable regression outcomes across scholar levels of political capital (see Table S3). For publication count, the funding coefficient is 2.847 (Politician 1) and 2.843 (Politician 2), revealing a strong positive impact of funding on publication output with near-identical effects across levels of academic political capital. Similarly, for citation count, the funding coefficients are 15.885 (Politician 1) and 15.918 (Politician 2), showing an almost identical significant positive effect. However, funding's effect on CiteScore remains smaller and insignificant, with coefficients of 0.473 for high-level political capital and 0.406 for lower-level political capital, neither reaching significance.

These results exhibit consistency in regressions for both high and low academic political capital and align with the direction of results when unclassified academic political capital is used as an instrumental variable. Whether classified or not, funding's significant positive impact on publication count and citation count is robustly supported, while its effect on CiteScore remains insignificant.

| Variables | Politician 1 | | | Politician 2 | | |
|---|---|---|---|---|---|---|
| | Publication | Citation | CiteScore | Publication | Citation | CiteScore |
| **Funding** | 2.847*** | 15.885*** | 0.473 | 2.843*** | 15.918*** | 0.406 |
| | (0.417) | (2.939) | (1.559) | (0.416) | (2.933) | (1.558) |
| **Initial ln (Article counts \| Citations \| Citescore)** | 0.620*** | 22.204*** | 0.831* | 0.619*** | 22.210*** | 0.818** |
| | (0.083) | (0.775) | (0.330) | (0.082) | (0.775) | (0.330) |
| **Funding quota** | -0.001 | 0.077 | 0.661*** | -0.001 | 0.077 | 0.660*** |
| | (0.013) | (0.177) | (0.079) | (0.013) | (0.177) | (0.079) |
| **Academic age** | 0.073*** | -0.258*** | 0.149*** | 0.073*** | -0.258*** | 0.149*** |
| | (0.003) | (0.027) | (0.014) | (0.003) | (0.027) | (0.014) |
| **Year** | 0.115*** | -0.098 | -0.291*** | 0.115*** | -0.097 | -0.292*** |
| | (0.007) | (0.085) | (0.040) | (0.007) | (0.085) | (0.040) |
| **Gender** | 0.278*** | 0.048 | 0.288 | 0.278*** | 0.047 | 0.290 |
| | (0.041) | (0.617) | (0.267) | (0.041) | (0.617) | (0.267) |
| **ln ( Publications $_{fields}$ )** | -0.376*** | -4.000*** | -5.531*** | -0.376*** | -3.999*** | -5.533*** |
| | (0.027) | (0.439) | (0.200) | (0.027) | (0.439) | (0.200) |
| **ln ( Publications $_{Affiliations}$ )** | 0.304*** | -4.782*** | -6.947*** | 0.303*** | -4.779*** | -6.953*** |
| | (0.069) | (0.674) | (0.298) | (0.069) | (0.674) | (0.298) |
| **ln ( Citations $_{fields}$ )** | 0.416*** | 4.199*** | 5.788*** | 0.417*** | 4.198*** | 5.790*** |
| | (0.028) | (0.445) | (0.209) | (0.028) | (0.445) | (0.209) |
| **ln ( Citations $_{Affiliations}$ )** | -0.286*** | 4.651*** | 6.725*** | -0.285*** | 4.648*** | 6.731*** |
| | (0.059) | (0.511) | (0.257) | (0.059) | (0.511) | (0.257) |
| **Employer reputation** | 0.002* | 0.050*** | 0.023*** | 0.002* | 0.050*** | 0.023*** |
| | (0.001) | (0.013) | (0.005) | (0.001) | (0.013) | (0.005) |
| **US NEWS ranking** | 0.000*** | 0.000 | 0.001*** | 0.000*** | 0.000 | 0.001*** |
| | (0.000) | (0.001) | (0.000) | (0.000) | (0.001) | (0.000) |
| **QS ranking** | 0 | 0.002** | -0.001** | 0.000 | 0.002** | -0.001** |
| | (0.000) | (0.001) | (0.000) | (0.000) | (0.001) | (0.000) |
| **Constant** | -232.676*** | 163.148 | 553.137*** | -232.543*** | 162.135 | 555.175*** |
| | (14.801) | (172.149) | (81.069) | (14.786) | (171.945) | (81.045) |

$p < 0.05$; ** $p < 0.01$; *** $p < 0.001$. The values in parentheses are standard errors.

**Table S3 2SLS regression results for academic publication metrics of two politicians.**

In summary, the multiple regression outcomes support the exogeneity of academic political capital as an instrumental variable. This finding substantiates the validity of the exogeneity assumption for scholars' political capital within this study. The similarity in results across different levels of academic political resources reflects that, while academic political positions may indirectly affect the distribution of academic resources, they do not directly impact academic output. Instead, academic success depends more substantially on individual productivity and institutional support than on accumulated political capital. This aligns with a "performance-oriented" view in academia, supporting the use of academic political capital as a stable instrumental variable to address the endogeneity of funding's effect on academic performance, thus providing robust theoretical and empirical support for explaining academic outcomes within this research framework (Muller, 2019).

## G. SHORT- AND LONG-TERM VALIDATION OF IVS

We use five years as the time window for the impact of the three IVs on researchers. The paper continues to test whether the IVs have a stable impact on scholars in the short and long term, i.e., whether the IVs remain valid when the impact stages of the IVs on scholars are assumed to be three and seven years, respectively.

Figure S4 shows our three IVs that are valid for measuring research funding and the three metrics, i.e., publication counts, citation counts, and CiteScore, respectively, under different sets of impact time windows. We employ the Hansen J test to check for exogeneity of IVs and the heteroskedasticity robust Durbin Wu-Hausman test (Baum et al., 2003) for endogeneity of SBE research funding. We see that there is a significant endogeneity issue in the process of measuring the SBE research funding effects on the science metrics. At the same time, IVs remain exogenous in the process of estimating various scientific metrics when the time windows for IVs are three and seven years.

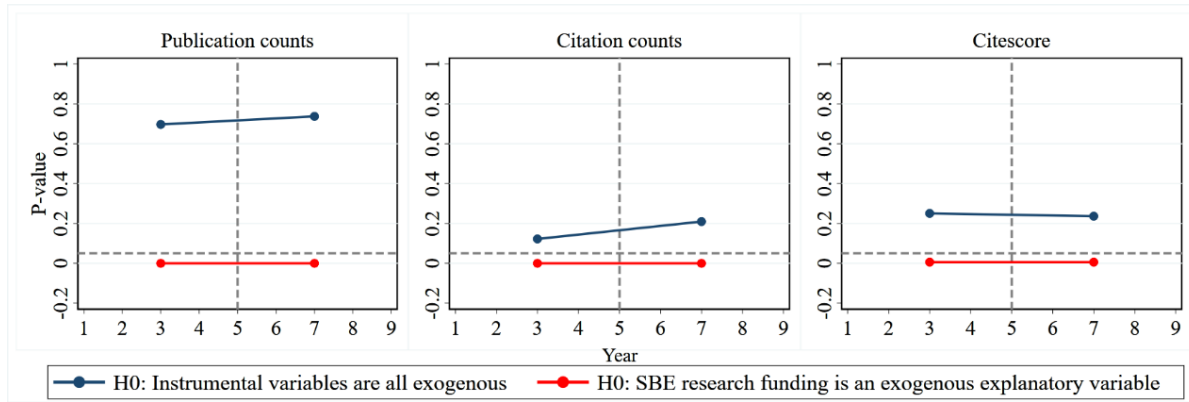

**Figure S4 Results of endogeneity tests for research grants and exogeneity tests for IVs.**

Further, we provide test results for these IVs in the first-stage regression of 2SLS under different time windows (see Table S4). We can observe that, when the time window is three or seven years, the *F*-statistics are all larger than 10 and 10% critical values of 2SLS size of nominal 5% Wald test and 5% maximal IV relative bias, implying that none of the IVs have zero coefficients in the first stage and are not weak IVs.

| | | **3 years** | **7 years** |
|---|---|---|---|
| **2SLS 1st stage** | Robust F | 546.078 | 544.253 |
| | Minimum eigenvalue statistic | 509.067 | 507.394 |
| | 2SLS size of nominal 5% Wald test (10% critical values) | 22.300 | |
| | 2SLS size of nominal 5% Wald test (15% critical values) | 12.830 | |
| | Stock-Yogo weak ID test critical values (5% maximal IV relative bias) | 13.910 | |

**Table S4 2SLS first-stage regression results for different time windows of IVs.**

## H. FURTHER DISCUSSION ON THE EXCLUSION RESTRICTION AND DESIGN OF THE IMITATION ISOMORPHISM INSTRUMENT

In this appendix, we first explore the establishment of the exclusion restriction for the imitation isomorphism instrument. We then provide a design steps for this instrument.

### H1 Background of Imitative Isomorphism as an Instrument

Our study utilizes an instrumental variable previously identified in the literature, focusing on the number of colleagues within the same affiliated institution who have secured funding (Kelchtermans et al., 2022). This instrument draws on the concept of "local conditions," positing that the funding success of peers can affect an individual scholar's own funding success. A critical assumption for the validity of this instrument is that funding granted to colleagues at the same institution is not expected to impact the research of other scientists. If funded researchers deploy their resources in a way that alters the research of their peers, the effectiveness of this instrument would be compromised, thus challenging the assumption of its exogeneity. We seek to clarify and redesign this instrument further. Specifically, we believe that merely controlling for the number of funded colleagues is insufficient; therefore, our research incorporates an additional dimension by considering the funding status of colleagues working on similar research topics within the same affiliation.

**H2 Confounding Factors Affecting the Effectiveness of Imitation Isomorphism**

Imitation isomorphism as an instrument may be susceptible to various confounding factors. For instance, a developed imitation isomorphism may indicate that the department has established a strong academic prestige, which in turn enhances the academic performance of other researchers. Let us denote $Z$ as imitation isomorphism, $X$ as research funding, $Y$ as academic performance metrics of scholars, and $U$ as confounding factors. Scholars $i$ and $j$ are affiliated with the same institution. We can illustrate the causal pathways of these confounding factors using a simplified causal diagram, as shown in Figure S5.

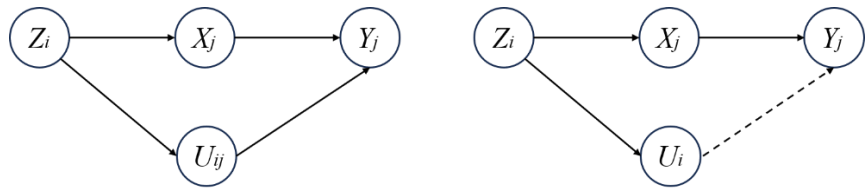

**Figure S5 Causal diagram for imitation isomorphism as an instrument.**

In the left sub-figure of Figure S5, the funding status of colleague $i$ may affect the prestige $U_{ij}$ of the affiliated institution where scholars $i$ and $j$ work. This confounder $U_{ij}$ could also directly affect the academic performance of scholar $j$. This causal pathway is valid if there is a connection between scholars $i$ and $j$. Being at the same institution facilitates this causal relationship, and the university's prestige $U_{ij}$ can interfere with the exogeneity assumption of the instrument $Z_i$. Therefore, it is essential to control for factors that can affect scholars $i$ and $j$ as beneficiaries of this shared prestige.

The right sub-figure of Figure S5 illustrates that the funded scholar $i$ may experience changes in some unobservable factors $U_i$ due to funding, such as enhanced academic connections, invitations to conferences, or dual appointments.

Unlike the left sub-figure, $Z_i$ affects $U_i$, while $Y_j$ refers to the academic performance of colleague $j$. Unlike $U_{ij}$ in the left sub-figure, which can affect $Y_j$, $U_i$ is unlikely to affect $Y_j$. However, unless there is a connection between $i$ and $j$ through research or institutional prestige, the $U_i$ factors affected by scholar $i$ are unlikely to impact the academic performance of scholar $j$.

To isolate the effect of scholars receiving funding on their colleagues' publishing productivity and impact via changes in university prestige, we control for their institutions' QS and US News rankings, employer reputation scores, as well as the number of publications and citation counts. This helps eliminate the causal pathway associated with institutional prestige, thus reducing potential confounding effects. Additionally, we exclude scholars $j$ who have co-authored publications with funded colleague $i$, further isolating potential confounding factor $U_{ij}$ and strengthening the exclusion restriction of the instrument. By removing scholars with collaborative networks, and controlling for their shared university prestige, it becomes less likely that the factors affecting scholar $i$ will impact the academic performance of a non-collaborating scholar $j$. This approach diverges from conventional analyses of instrumental variables, as in this scenario, unobservable factors tied to specific individual interests are unlikely to affect the performance of another distinct individual. The benefits $U_i$ derived from research funding for scholar $i$ will not affect $Y_j$, thus alleviating the need for further isolation.

Furthermore, our robustness checks regarding the instrumental variable's stability, and its long- and short-term effectiveness, further validate the credibility of imitative isomorphism as an instrument.

### H3 Steps in Constructing Imitation Isomorphism

The second IV is operationalized as follows.

*Step 1* The LDA model is adopted to extract research topics from the proposal titles and abstracts of all recipients, resulting in a total of 30 broad categories of research topics.

*Step 2* We count the number of grants won by all universities in different research topics over the years, as shown in Figure S6.

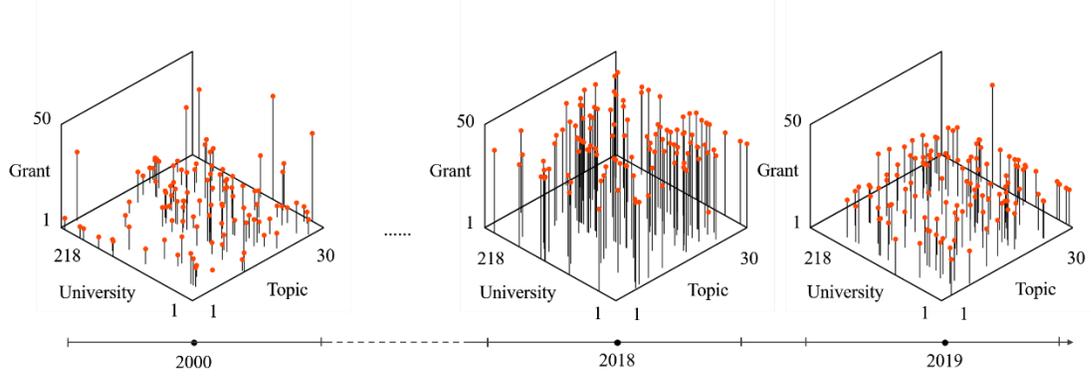

**Figure S6 Distribution of the number of grants received by universities on different research topics per year.**

*Step 3* We cumulatively sum the number of awards the university receives each year for different topics and take the reverse value. Let us take the award record of the University of Michigan at Ann Arbor as an example. After cumulative awards, its cumulative number of awards in Topic 3: 'Humans, Species, Genetics, Evolution, Data' is 15 in 2019 and 0 in 2000. We then take the reverse value of this cumulative and time series, i.e., Michigan's award difficulty or the reverse value of imitation isomorphism in 2000 is 15; then, the award difficulty begins to decrease.

*Step 4* We normalize the number of funded projects for each university's research topic to balance the order of magnitude difference in the number of projects. The imitation isomorphism (II) is calculated as follows

$$II^{reversed} = \frac{II - min(II)}{max(II) - min(II)} \tag{8}$$

where $II$ represents the reverse of the cumulative number of awards in Step 3. Similarly, take the case of Step 3 as an example. Again, using Michigan as an example, its reverse value of 15 in 2000 represents $max(II)$, which is assigned a weight of 1; its reverse value of 5 in 2019 represents $min(II)$, which is weighted at 0.120.

## I. ANALYSIS OF THE EFFECTS OF RESEARCH FUNDING ON TOP 10% AND BOTTOM 10% CITESCORE JOURNALS

To investigate the heterogeneous effects of research funding across different levels of journal impact, we created two new dependent variables based on journal CiteScore rankings: 1) Top 10% CiteScore Journals: For this variable, we retained the original CiteScore for publications in the top 10% of the CiteScore distribution. All other values were replaced with the average CiteScore of the scholars' publications. 2) Bottom 10% CiteScore Journals: For this variable, we retained the original CiteScore for publications in the bottom 10% of the CiteScore distribution. All other values were replaced with the average CiteScore of the scholars' publications. This design allows us to isolate the impact of research funding on publications in the highest and lowest impact journals while controlling for the overall publication visibility/impact of each scholar. The approach provides a focused lens to evaluate whether funding disproportionately benefits scholars aiming to publish in lower-tier journals or helps them access more prestigious outlets.

The decision to focus on the top and bottom 10% was motivated by the need to understand whether research funding plays a *redistributive* role in academic outcomes. Specifically, does funding elevate the baseline performance of scholars struggling to publish in high-impact outlets, or does it enhance their ability to compete for space in the most prestigious journals? Identifying these effects is crucial for understanding the systemic impact of funding policies on academic inequalities and the broader research ecosystem.

## I1 Results

Table S5 summarizes the regression results for the Top 10% and Bottom 10% CiteScore journals. The findings highlight a clear asymmetry in the effects of funding. Research funding does not have a statistically significant effect on scholars' ability to publish in the most prestigious journals (Top 10% CiteScore Journals, coef.=−0.650, p=0.547). This suggests that success in top-tier outlets is driven by factors such as institutional reputation and pre-existing academic achievements, where additional funding offers limited incremental benefits.

| Variable | Top 10% Coefficient (Std. Err.) | Bottom 10% Coefficient (Std. Err.) |
|---|---|---|
| **Funding** | -0.650 (1.079) | 3.183*** (1.039) |
| **Initial Productivity** | 0.163 (0.232) | 1.107*** (0.211) |
| **Initial ln (Article counts | Citations | Citescore)** | 0.626*** (0.054) | -0.056 (0.046) |
| **Career** | 0.001 (0.008) | 0.175*** (0.010) |
| **Funding quota** | -0.098*** (0.029) | -0.137*** (0.022) |
| **Gender** | 0.277 (0.186) | -0.081*** (0.151) |
| **Academic age** | -2.166*** (0.142) | -2.282*** (0.109) |
| **ln(Article Affiliations)** | -3.255*** (0.210) | -1.718*** (0.184) |
| **Year** | 2.255*** (0.148) | 2.410*** (0.114) |
| **ln(Citation Affiliations)** | 3.240*** (0.179) | 1.685*** (0.165) |
| **Gender** | 0.030*** (0.004) | -0.011*** (0.003) |
| **USNEWS Ranking** | 0.0003 (0.0002) | 0.0006*** (0.0002) |
| **ln ( Publications $_{fields}$ )** | 0.0002 (0.0002) | -0.0009*** (0.0002) |
| **Constant** | 205.441*** (57.917) | 274.336*** (45.353) |

* p < 0.05; ** p < 0.01; *** p < 0.001. The values in parentheses are standard errors.

**Table S5** Regression results of research funding on top 10% and bottom 10% journals' CiteScore.

In contrast, funding significantly improves the impact of publications in the bottom 10% CiteScore journals (coef.=3.183, p=0.002). This indicates that funding helps scholars improve the quality of their work in lower-tier journals, thereby elevating their baseline academic performance.

## I2 Instrumental variable validity

Table S6 presents key diagnostic results for the validity and robustness of the instrumental variables used in the analysis. The Kleibergen-Paap rk LM statistic confirms that there is no under-identification issue in either the Top 10% or Bottom 10% regression models (p=0.000), indicating that the instrumental variables are sufficiently relevant for explaining the allocation of research funding. Additionally, the Kleibergen-Paap rk Wald F statistic exceeds the critical values provided by the Stock-Yogo weak instrument test, demonstrating strong instrument relevance and ruling out weak instrument concerns. The Hansen J statistic further validates the instruments, showing no significant correlation with the error term (Top 10% regression p=0.215; Bottom 10% regression p=0.246), thereby supporting the overidentification restrictions. Moreover, the endogeneity tests reveal that research funding is endogenous in both the Top 10% model (p=0.040) and the Bottom 10% model (p=0.003), highlighting the necessity of addressing endogeneity in these analyses. Taken together, these findings confirm that the chosen instrumental variables are both valid and robust in addressing potential endogeneity bias, ensuring the reliability and interpretability of the regression estimates.

| Test | Top 10% Results | Bottom 10% Results |
|---|---|---|
| Underidentification Test (Kleibergen-Paap rk LM statistic) | 1278.574 (p=0.000) | 1278.613 (p=0.000) |
| Weak Identification Test (Kleibergen-Paap rk Wald F statistic) | 543.782 | 543.802 |
| Stock-Yogo Critical Value (10% Max IV Size) | 22.300 | 22.300 |
| Hansen J Statistic (Overidentification Test) | 3.074 (p=0.215) | 2.806 (p=0.246) |
| Endogeneity Test for Funding | p=0.040) | p=0.003 |

The critical values for the Cragg-Donald Wald F statistic are based on Hausman et al. (2005).

**Table S6** Diagnostic tests for instrumental variable validity and endogeneity.

### I3 Discussion

The results provide important insights into the role of research funding in shaping academic outcomes across journals with varying impact levels. While funding does not significantly enhance the likelihood of publishing in top-tier journals, it plays a pivotal role in raising the quality and impact of publications in lower-tier outlets. This suggests that research funding is particularly effective in addressing resource constraints and improving the academic performance of scholars operating at the lower end of the journal impact spectrum. These findings have significant implications for research funding policies. By lifting the "floor" of academic performance, funding can help reduce inequalities in scholarly outcomes and foster a more inclusive research environment. However, the lack of impact on the "ceiling" indicates that additional strategies may be required to support scholars in competing for space in the most prestigious journals. Overall, these results underscore the importance of designing funding mechanisms that balance short-term productivity goals with long-term research quality and innovation.

## J. ROBUSTNESS TEST RESULTS FOR SBE FUNDING EFFECTS

When controlling for endogeneity and potential bias in research funding, there should be no significant difference in funding effects between the two groups of academics receiving research funding at the same time; there should also be no research funding effects between academics who have not yet received funding. We randomly assign research grants in each of the two groups of individuals and observe the research grant effect between individuals in the treated and control groups separately, repeating the 2SLS strategy.

Tables S7 and S8 demonstrate that there is no research funding effect between individuals in the treated and control groups, respectively. It is assumed that randomly labelled funded individuals are in a non-funded status, i.e., a pseudo-control group; control individuals receiving pseudo-research funding are referred to as the pseudo-treated group. The funding effects measured by both article counts, citation counts, and CiteScore are not significant for individuals in the treated and pseudo-control groups (Table S7) and for individuals in the control and pseudo-treated groups (Table S8). This is one piece of evidence that implies that our strategy is effective in isolating the endogeneity issue and demonstrates the reliability of research funding effects among individuals in the original treated group and the original control group.

| | 2SLS 1$^{st}$ stage | 2SLS 2$^{nd}$ stage |
|---|---|---|

| Variables | Funding | Publication counts | Citation counts | CiteScore |
|---|---|---|---|---|
| **Funding** | - | -3.568<br>(1.932)<br>($p$=0.079) | -60.216<br>(35.522)<br>($p$=0.090) | -4.108<br>(10.699)<br>($p$=0.677) |
| **Initial ln (Publicationcounts \| Citations \| CiteScore)** | -0.175<br>(0.001)<br>($p$=0.000) | -0.56<br>(0.339)<br>($p$=0.119) | 3.039<br>(6.129)<br>($p$=0.620) | 0.087<br>(1.866)<br>($p$=0.987) |
| **Funding quota** | 0.000<br>(0.001)<br>($p$=0.903) | -0.025<br>(0.016)<br>($p$=0.108) | 0.356<br>(0.383)<br>($p$=0.352) | 0.82<br>(0.094)<br>($p$=0.000) |
| **Academic age** | 0.003<br>(0.000)<br>($p$=0.000) | 0.078<br>(0.008)<br>($p$=0.000) | -0.068<br>(0.135)<br>($p$=0.615) | 0.161<br>(0.039)<br>($p$=0.000) |
| **Year** | -0.004<br>(0.000)<br>($p$=0.000) | 0.078<br>(0.009)<br>($p$=0.000) | -2.830<br>(0.176)<br>($p$=0.000) | -0.284<br>(0.051)<br>($p$=0.000) |
| **Gender** | -0.017<br>(0.004)<br>($p$=0.000) | 0.407<br>(0.061)<br>($p$=0.000) | -1.595<br>(1.480)<br>($p$=0.281) | -0.240<br>(0.373)<br>($p$=0.510) |
| **ln ( Publications$_{fields}$ )** | -0.036<br>(0.003)<br>($p$=0.000) | -0.621<br>(0.076)<br>($p$=0.000) | -14.278<br>(1.508)<br>($p$=0.000) | -6.010<br>(0.446)<br>($p$=0.000) |
| **ln ( Publications$_{Affiliations}$ )** | -0.148<br>(0.005)<br>($p$=0.000) | -0.393<br>(0.291)<br>($p$=0.215) | -23.313<br>(5.175)<br>($p$=0.000) | -8.067<br>(1.713)<br>($p$=0.000) |
| **ln ( Citations$_{fields}$ )** | 0.036<br>(0.003)<br>($p$=0.000) | 0.660<br>(0.078)<br>($p$=0.000) | 14.43<br>(1.492)<br>($p$=0.000) | 6.285<br>(0.452)<br>($p$=0.000) |
| **ln ( Citations$_{Affiliations}$ )** | 0.143<br>(0.005)<br>($p$=0.000) | 0.382<br>(0.281)<br>($p$=0.211) | 22.227<br>(4.989)<br>($p$=0.000) | 7.830<br>(1.636)<br>($p$=0.000) |
| **Employer reputation** | 0.000<br>(0.000)<br>($p$=0.000) | 0.001<br>(0.001)<br>($p$=0.385) | 0.081<br>(0.024)<br>($p$=0.001) | 0.021<br>(0.007)<br>($p$=0.006) |
| **US NEWS ranking** | 0.000<br>(0.000)<br>($p$=0.166) | 0.001<br>(0.000)<br>($p$=0.001) | 0.001<br>(0.001)<br>($p$=0.151) | 0.001<br>(0.000)<br>($p$=0.005) |
| **QS ranking** | 0.001<br>(0.000)<br>($p$=0.000) | 0.001<br>(0.000)<br>($p$=0.035) | 0.001<br>(0.002)<br>($p$=0.803) | -0.001<br>(0.001)<br>($p$=0.100) |
| **Isomorphism** | 0.000<br>(0.000)<br>($p$=0.001) | | | |
| **Political hegemony** | 0.078<br>(0.028)<br>($p$=0.005) | | | |
| **NSF training** | -0.006<br>(0.001)<br>($p$=0.000) | | | |
| **Constant** | 7.460 | -155.186 | 5641.263 | 537.572 |

|  | (0.905) | (17.671) | (357.503) | (104.272) |
|  | (*p*=0.000) | (*p*=0.000) | (*p*=0.000) | (*p*=0.000) |

**Table S7 Results of the funding effect between individuals in the treated and pseudo-control groups.**

| Variables | 2SLS 1st stage | 2SLS 2nd stage | | |
|---|---|---|---|---|
|  | Funding | Publication counts | Citation counts | CiteScore |
| **Funding** |  | 0.211 | 6.875 | 3.547 |
|  |  | (0.109) | (5.745) | (1.887) |
|  |  | (*p*=0.053) | (*p*=0.231) | (*p*=0.060) |
| **Initial ln (Publication counts \| Citations \| CiteScore)** | -0.153 | 0.109 | 16.752 | 2.237 |
|  | (0.003) | (0.020) | (1.436) | (0.381) |
|  | (*p*=0.000) | (*p*=0.000) | (*p*=0.000) | (*p*=0.000) |
| **Funding quota** | -0.007 | 0.007 | -0.508 | 0.424 |
|  | (0.002) | (0.007) | (0.279) | (0.146) |
|  | (*p*=0.000) | (*p*=0.309) | (*p*=0.069) | (*p*=0.004) |
| **Academic age** | 0.002 | 0.010 | -0.480 | 0.223 |
|  | (0.000) | (0.002) | (0.069) | (0.034) |
|  | (*p*=0.000) | (*p*=0.000) | (*p*=0.000) | (*p*=0.000) |
| **Year** | 0.045 | 0.052 | 0.583 | 0.478 |
|  | (0.001) | (0.006) | (0.369) | (0.108) |
|  | (*p*=0.000) | (*p*=0.000) | (*p*=0.114) | (*p*=0.000) |
| **Gender** | -0.001 | -0.039 | 2.539 | 1.063 |
|  | (0.007) | (0.022) | (0.965) | (0.449) |
|  | (*p*=0.899) | (*p*=0.071) | (*p*=0.009) | (*p*=0.018) |
| **ln ( Publications $_{fields}$ )** | -0.022 | -0.155 | -6.128 | -4.533 |
|  | (0.005) | (0.017) | (1.006) | (0.373) |
|  | (*p*=0.000) | (*p*=0.000) | (*p*=0.000) | (*p*=0.000) |
| **ln ( Publications $_{Affiliations}$ )** | -0.084 | -0.030 | -6.431 | -4.669 |
|  | (0.008) | (0.024) | (1.783) | (0.494) |
|  | (*p*=0.000) | (*p*=0.219) | (*p*=0.000) | (*p*=0.000) |
| **ln ( Citations $_{fields}$ )** | 0.025 | 0.160 | 6.695 | 4.715 |
|  | (0.005) | (0.018) | (1.063) | (0.395) |
|  | (*p*=0.000) | (*p*=0.000) | (*p*=0.000) | (*p*=0.000) |
| **ln ( Citations $_{Affiliations}$ )** | 0.071 | 0.009 | 5.732 | 4.451 |
|  | (0.007) | (0.020) | (1.211) | (0.386) |
|  | (*p*=0.000) | (*p*=0.668) | (*p*=0.000) | (*p*=0.000) |
| **Employer reputation** | 0.001 | 0.001 | 0.063 | 0.034 |
|  | (0.000) | (0.000) | (0.021) | (0.009) |
|  | (*p*=0.860) | (*p*=0.781) | (*p*=0.002) | (*p*=0.000) |
| **US NEWS ranking** | 0.001 | 0.001 | -0.001 | 0.001 |
|  | (0.000) | (0.000) | (0.001) | (0.000) |
|  | (*p*=0.648) | (*p*=0.001) | (*p*=0.410) | (*p*=0.085) |
| **QS ranking** | -0.001 | -0.000 | 0.002 | -0.001 |
|  | (0.000) | (0.000) | (0.001) | (0.000) |
|  | (*p*=0.012) | (*p*=0.698) | (*p*=0.093) | (*p*=0.398) |
| **Isomorphism** | 0.003 |  |  |  |
|  | (0.001) |  |  |  |
|  | (*p*=0.000) |  |  |  |
| **Political hegemony** | 0.151 |  |  |  |
|  | (0.040) |  |  |  |
|  | (*p*=0.000) |  |  |  |
| **NSF training** | -0.008 |  |  |  |

|  |  |  |  |
|---|---|---|---|
| (0.002) | | | |
| ($p$=0.000) | | | |

| | Constant | -90.472 | -104.136 | -1199.210 | -988.687 |
|---|---|---|---|---|---|
| | | (2.822) | (11.723) | (739.322) | (217.522) |
| | | ($p$=0.000) | ($p$=0.000) | ($p$=0.105) | ($p$=0.000) |

**Table S8 Results of the funding effect between individuals in the control and pseudo-treated groups.**

Further, we provide the validity tests for the IVs in the two robustness tests, respectively (Tables S9 and S10). It is noticeable that all IVs remain valid in the subgroups of treated and pseudo-control groups, and control and pseudo-treated groups. This implies that, even if we randomly assign scholars to receive research grants, the serious endogeneity of research grants is not offset by the effect of our randomization within the group of scholars. The endogeneity issue of research grants in the SBE still requires IVs to be effectively isolated. The test results also provide a side-effect explanation for the efficacy of our choice of three IVs in isolating endogeneity.

| | | | | Publication counts | Citation counts | CiteScore |
|---|---|---|---|---|---|---|
| **(1)** | **SBE funding endogeneity** | **Hausman test** | **Robust score chi2(1)** | 5.793 ($p$=0.016) | 8.430 ($p$=0.004) | 6.741 ($p$=0.009) |
| | | | **Robust regression $F$** | 5.813 ($p$=0.016) | 8.410 ($p$=0.004) | 6.765 ($p$=0.009) |
| **(2)** | **Over -identification** | **Hansen J test** | **Statistics** | 5.110 ($p$=0.078) | 2.241 ($p$=0.326) | 0.323 ($p$=0.851) |
| **(3)** | **Under -identification** | **Kleibergen-Paap rk LM test** | **Chi-sq** | 45.315 ($p$=0.000) | | |
| **(4)** | **Weak identification test** | **Cragg-Donald Wald test** | **F statistic** | 16.311 | | |
| | | **Stock-Yogo weak ID test** | **5% maximal IV relative bias** | 12.830 | | |
| **(5)** | **2SLS first stage** | **Robust F** | | 15.672 ($p$=0.000) | | |
| | | **Minimum Eigenvalue Statistics** | | 16.814 | | |
| | | **2SLS size of nominal 5% Wald test (10% critical values)** | | 22.300 | | |
| | | **2SLS size of nominal 5% Wald test (15% critical values)** | | 12.830 | | |

**Table S9 Results of validity and robustness tests of IVs (treated group).**

| | | | Publication counts | Citation counts | CiteScore |
|---|---|---|---|---|---|
| (1) | SBE funding endogeneity | Hausman test | Robust score chi2(1) | 10.644 ($p$=0.001) | 19.222 ($p$=0.000) | 4.687 ($p$=0.030) |
| | | | Robust regression $F$ | 10.627 ($p$=0.001) | 19.301($p$=0.000) | 0.972 ($p$=0.009) |
| (2) | Over-identification | Hansen J test | Statistics | 4.993 ($p$=0.082) | 0.026 ($p$=0.987) | 0.323 ($p$=0.615) |
| (3) | Under-identification | Kleibergen-Paap rk LM test | Chi-sq | 374.813 ($p$=0.000) | | |
| (4) | Weak identification test | Cragg-Donald Wald test | F statistic | 272.180 | | |
| | | Stock-Yogo weak ID test | 5% maximal IV relative bias | 13.910 | | |
| (5) | 2SLS first stage | Robust F | | 231.879 ($p$=0.000) | | |
| | | Minimum Eigenvalue Statistics | | 272.180 | | |
| | | 2SLS size of nominal 5% Wald test (10% critical values) | | 22.300 | | |
| | | 2SLS size of nominal 5% Wald test (15% critical values) | | 12.830 | | |

**Table S10 Results of validity and robustness tests of IVs (control group).**

## SUPPLEMENTARY REFERENCES